\pdfoutput=1

\documentclass[useAMS,usenatbib]{mn2e}

\usepackage{txfonts}
\usepackage{xspace}
\usepackage{graphicx}
\usepackage[usenames,dvipsnames]{color}

\newcommand{\Lsun}{\ensuremath{\mathrm{L}_{\odot}}}
\newcommand{\iprn}{\mbox{$i$ process}}
\newcommand{\ipr}{\mbox{$i$-process}}
\newcommand{\sprn}{\mbox{$s$ process}}
\newcommand{\spr}{\mbox{$s$-process}}
\newcommand{\rprn}{\mbox{$r$ process}}
\newcommand{\rpr}{\mbox{$r$-process}}
\newcommand{\nuclei}[2]{\ensuremath{\mathrm{^{#1}#2}}}
\newcommand{\carbon}[1][12]{\nuclei{#1}{C}}
\newcommand{\ocen}{\mbox{$\Omega$ Cen}}
\newcommand{\oxygen}[1][16]{\nuclei{#1}{O}}
\newcommand{\neon}[1][20]{\nuclei{#1}{Ne}}
\newcommand{\nt}{\neutron}
\newcommand{\neutron}{\ensuremath{n}}
\newcommand{\magnesium}[1][24]{\nuclei{#1}{Mg}}
\newcommand{\mem}[1]{\ensuremath{\mathrm{ #1}}}
\newcommand{\Sectff}[1]{{\ref{sec:#1}}}
\newcommand{\silicon}[1][28]{\nuclei{#1}{Si}}
\newcommand{\nitrogen}[1][14]{\nuclei{#1}{N}}
\newcommand{\helium}[1][4]{\nuclei{#1}{He}}
\newcommand{\natlog}[2]{\ensuremath{#1\times 10^{#2}}} 
\newcommand{\proton}{\ensuremath{p}}
\newcommand{\power}[2]{\ensuremath{{#1}^{#2}}}
\newcommand{\Sect}[1]{{\S\Sectff{#1}}}

\newcommand{\msun}{\ensuremath{\,M_\odot}}			
\newcommand{\fenv}{\ensuremath{f_{\rm env}}}               
\newcommand{\fPDCZ}{\ensuremath{f_{\rm PDCZ}}}                  

\newcommand{\mdot}{\ensuremath{\dot{M}}}

\title[Hydrogen ingestion in super-AGB stars]{H ingestion into He-burning convection zones in super-AGB stellar models as a potential site for
intermediate neutron-density nucleosynthesis}
\author[S.~W.~Jones et al.]{S.~Jones$^{1,2,\dagger}$\thanks{Email: samuel.jones@h-its.org}, 
C.~Ritter$^{1,3,\dagger}$,
F.~Herwig$^{1,3,\dagger}$,
C.~Fryer$^{4,\dagger}$, M.~Pignatari$^{5,\dagger}$,
M.~G.~Bertolli$^{6,7,\dagger,\ddagger}$
\newauthor
and B.~Paxton$^8$ \\
$^1$ Department of Physics \& Astronomy, University of Victoria, P.O. Bos 3055 Victoria, B.C., V8W 3P6, Canada \\
$^2$ Heidelberg Institute for Theoretical Studies, Schloss-Wolfsbrunnenweg 35, D-69118 Heidelberg, Germany \\
$^3$ Joint Institute for Nuclear Astrophysics, Center for the Evolution of the Elements, Michigan
State University \\ 640 South Shaw Lane, East Lansing, MI 48824, USA \\
$^4$ Computational Physics and Methods (CCS-2), LANL, Los Alamos, NM, 87545, USA \\
$^5$ Konkoly Observatory, Research Centre for Astronomy and Earth Sciences, Hungarian Academy of Sciences, \\ Konkoly-Thege Mikl\'{o}s ut 15-17, H-1121 Budapest, Hungary \\
$^6$ Physics Division, Oak Ridge National Laboratory, Oak Ridge, Tennessee 37831, USA. \\
$^7$ Department of Physics and Astronomy, University of Tennessee, Knoxville, Tennessee 37996, USA \\
$^8$ Kavli Institute for Theoretical Physics and Department of Physics Kohn Hall,
University of California, Santa Barbara, CA 93106, USA \\
$^\dagger$ NuGrid Collaboration, \emph{http://www.nugridstars.org} \\
$^\ddagger$ Formerly LANL}
\begin{document}
\date{Submitted: June 2015}
\pagerange{\pageref{firstpage}--\pageref{lastpage}} \pubyear{2015}
\maketitle
\label{firstpage}

\begin{abstract}
We investigate the evolution of super-AGB thermal pulse (TP) stars for a
range of metallicities ($Z$) and explore the effect of
convective boundary mixing (CBM).
With decreasing metallicity and evolution along the TP phase,
the He-shell flash and the third dredge-up (TDU)
occur closer together in time.
After some time (depending upon the CBM parameterisation),
efficient TDU begins while the pulse-driven convection zone (PDCZ) is still present,
causing a convective exchange of material between the PDCZ and
the convective envelope.
This results in the ingestion of protons into the convective He-burning pulse.
Even small amounts of CBM encourage the interaction of the convection
zones leading to transport of protons
from the convective envelope into the He layer.
H-burning luminosities exceed $10^9$
(in some cases $10^{10})~\Lsun$.
We also calculate models of dredge-out in the most massive super-AGB stars
and show that the dredge-out phenomenon is another
likely site of convective-reactive H-\carbon[12] combustion.
We discuss the substantial uncertainties of
stellar evolution models under these
conditions.
Nevertheless, the simulations suggest
that in the convective-reactive H-combustion regime of H-ingestion
the star may encounter conditions for the intermediate neutron
capture process (\iprn).
We speculate that some CEMP-s/r stars could originate in i-process
conditions in the
H-ingestion phases of low-$Z$ SAGB stars.
This scenario would however suggest a very low electron-capture supernova
rate from super-AGB stars.
We also simulate potential outbursts triggered by such H-ingestion events, present
their light curves and briefly discuss their transient properties.
  
\end{abstract}

\section{Introduction}
\label{s.introduction}
The most important mixing process in stars is convection. One of the
least-understood aspects of stellar convection is the convective
boundary. The way in which convective boundaries are modelled
determines many global properties that characterise the evolution
of stellar models. In
the most simplistic view the convective boundary is given by the
Schwarzschild condition ($\nabla_\mathrm{rad}>\nabla_\mathrm{ad}$).
However, the literal implementation of the
Schwarzschild boundary would imply a spherically symmetric
composition
discontinuity that does not exist in real stars, as shown by
hydrodynamic simulations and observations
\citep[see, e.g.][]{freytag:96,Schroder:1997uh,deupree:00,Ribas:2000fn,
herwig:06a,rogers:06,Meakin:2007dj,Herwig:2007wh,woodward:08a,
mocak:09,tian:09,mocak:11,Tkachenko:2014fp}.
Numerous strategies have been employed to model the properties of convective
boundaries.

Here we use the simple model of exponentially decaying
convective boundary mixing (CBM) initially based on hydrodynamic
simulations by \citet{freytag:96} and first introduced in
stellar evolution simulations by \citet{herwig:97} to
address the observational properties of H-deficient post-AGB stars
\citep{werner:06}. This or similar exponential models for
CBM have been adopted and investigated for many
other problems, including the \carbon[13]-pocket in low-mass AGB
stars \citep{Lugaro2003,Herwig:2003du,Cristallo:2009cu}, the
observed CNO enhancement of nova ejecta and their fast rise time
\citep{Denissenkov2013} and the quenching of the C flame in
super-AGB stars \citep{Denissenkov:2013dd} and the general
properties of AGB stellar models
\citep{Herwig:2000ua,MillerBertolami:2006dr,weiss:09}. A
quantitative determination of the CBM efficiency in terms of the
e-folding distance of the diffusion coefficient by analysis of
hydrodynamic simulations has been attempted
by \cite{Herwig:2007wh}, but the results obtained so far are preliminary
because they are based on 2D calculations.

The evolution of super-AGB stars is characterised by the ignition of
C burning that will lead to the formation of an
electron-degenerate ONe core 
\citep{garcia-berro:94}.
The details of the outcome of C burning has
been shown to depend on the assumptions for convective
boundaries. \citet{Denissenkov:2013dd}, \citet{Chen2014} and \citet{Farmer2015}
have demonstrated that in
models in which convective boundary mixing is taken into account the
C-flame is quenched and hybrid CONe WDs may form, with numerous
possible consequences that impact, for example, upon the progenitor evolution of
type Ia supernovae \citep{Denissenkov2015,Kromer2015}.

Further burning stages beyond C burning (i.e. Ne, O, Si) are prevented by a
combination of electron degeneracy and neutrino cooling.
Super-AGB stars may end their lives as
ONe (or hybrid CONe; \citealp{Denissenkov:2013dd}) white dwarfs or as
electron-capture core collapse supernovae
\citep{Nomoto1984,Nomoto1987,poelarends:08,Jones2013,Doherty2015}. 
The evolution up to the thermally pulsing super-AGB phase is characterized
by several dredge-up (DUP) events, including, in some
cases, a dredge-out event that brings larger amounts of He-burning
products to the surface \citep[e.g.][]{ritossa:99,siess:07,Herwig2012mltalf}.
Since the
evolution of the H-shell depends on the abundance of (C+N+O), super-AGB
thermal pulse stars that have enhanced C and O abundances because of
this dredge-out will afterwards behave as if they had larger initial
C+N+O abundances.

The initial mass range for entering the super-AGB phase is
rather narrow, from $\approx 7.5 - 9.25~\msun$ at solar metallicity according
to \citet{poelarends:08}, $8-9.75~\msun$ according to \citet{Doherty2015} and
$7-9~\msun$ according to \citet{Woosley2015}. This result is uncertain as it depends
sensitively on the treatment of the H- and the He-core convection
boundary.
The exact mass
range will also depend on the uncertain $\carbon + \carbon$ nuclear reaction
rate.
The limiting initial mass for C ignition decreases
from about 7.1~\msun\ to about 5.4~\msun\
when the C-burning rate is increased from $0.1\lambda_\mathrm{CF88}$
to $1000\lambda_\mathrm{CF88}$ \citep{Chen:2014gsa}, where
$\lambda_\mathrm{CF88}$ is the rate from \citet{caughlan:88}. This corresponds
to a decrease in the limiting core mass from 1.1~\msun\ to 0.93~\msun.
At lower metallicity the mass range
for the formation of super-AGB stars will span lower initial masses
\citep[see, e.g.,][]{siess:07}.

The composition of the wind ejecta and evolutionary fate of super-AGB
stars at low metallicity is of particular importance for the
interpretation of multiple populations observed in some globular
clusters, in particular the He-rich blue main-sequence in
\ocen\ \citep{anderson:97} and several other clusters.
These second (or third) generation populations have also unusual C, N and O
abundances and models of super-AGB stars
\citep[e.g.][]{ventura:11,Herwig2012mltalf,Karakas2014} have
been used to address the abundance anomalies of distinct populations in
globular clusters. \citet{Herwig2012mltalf} have proposed the
\emph{Galactic plane passage gas purging chemical evolution scenario} to
integrate photometric and stellar evolution information with dynamic
constraints from the cluster's orbit to address many of the observed
abundance properties of the distinct populations in \ocen.
In this work
we are reporting on further, and more detailed investigations of the
super-AGB phase in stellar models.

Different groups using different stellar evolution codes obtain
reasonably similar results for the super-AGB evolution if the same
assumptions (in particular for mixing) are made
\citep{poelarends:08,Siess2010,doherty:10,Doherty2014}. 
While some of these studies go beyond the strict Schwarzschild
criterion for convection, none of these previous investigations have, to
our knowledge, explored the effect of variable CBM efficiencies.

The effect of CBM in low-mass AGB stars has
been investigated in detail via simple CBM
algorithms which effectively model a mixing diffusion coefficient that
exponentially decreases with geometric distance from the convection boundary
\citep{herwig:97,mazzitelli:99,Herwig:2000ua,MillerBertolami:2006dr,weiss:09}.
The consequences of CBM for He-shell flash
convection are larger \carbon\ and \oxygen\ abundances in the intershell, in
agreement with observations of H-deficient post-AGB stars
\citep{werner:06}, and larger peak temperatures for the
\neon[22]-fuelled \spr\ contribution. \citet{Lugaro2003} have argued that
this may lead to isotopic ratio predictions, for example for Zr, that
could be incompatible with observations from pre-solar grains. This
question needs to be revisited however, considering possible modifications to
the details of the mixing algorithm as well as the latest nuclear
reaction rate data, in particular for the $\neon[22](\alpha,\nt)\magnesium[25]$
neutron source reaction rate \citep[e.g.][]{Longland2012,Bisterzo2015} and Zr
neutron capture rates \citep{Lugaro2014,Liu2014}.
\citet{Denissenkov2013} have shown that with exponential-diffusive CBM
the observed enhancement of C and O in nova ejecta and the fast rise time
of the nova light curve can be reproduced in 1D models. CBM has been reported
in multidimensional simulations of nova shell flashes \citep{casanova:11,Glasner2012}.
It is also accepted that CBM in the form of some sort of
overshooting prescription needs to be employed during the core
H-burning phase on the main sequence, and almost all stellar evolution
calculations do
account for this convection-induced extra mixing.

Rotationally-induced mixing at the bottom of the convection zone
has been shown to be insufficient in spherically-symmetric simulations
\citep{herwig:03b,siess:04} to model
the partial mixing between the H-rich envelope and the \carbon-rich core
that leads to the formation of the \carbon[13]-pocket in low-mass AGB stars
\citep[e.g.][]{Herwig:2000ua,cristallo:09a}. In addition,
it may reduce significantly the
\spr\ yields \citep[][]{piersanti:13}. How well rotationally-induced mixing can
be approximated in one-dimensional models of stars is unclear at present.

Thus, there is significant motivation to investigate how the
evolution of super-AGB stars proceeds if small and moderate
efficiencies of CBM are taken into account at all convective
boundaries.  In the case of super-AGB stars, this makes the
computations numerically much more demanding, and in many cases the
quantitative details of the results have to be considered as very
uncertain. This is so because (i) the exact amount and efficiency of
CBM is presently unknown for the convection boundaries in super-AGB
stars, and (ii) we frequently encounter in these calculations of
convective-reactive H-combustion regimes that at least locally violate some of the
assumptions that are the basis of 1D spherically symmetric stellar
evolution codes with the mixing-length theory for convection. In that sense, the
computations presented here are in a way to be considered as a road
map which points out areas in which more detailed 3D hydrodynamic
investigations need to be performed in the future.

In this paper we report on the occurrence of convective-reactive
H-combustion flashes in super-AGB thermal pulses in
stellar models when CBM is included. These H-ingestion flashes present
specific differences from H-ingestion flashes found previously in normal AGB stars of very low metallicity
\citep[e.g.][]{fujimoto:00,herwig:03a,suda:04,campbell:08,Cristallo:2009cu}. In
particular, the latter are usually a one-off event which removes,
through copious dredge-up of C and O, the conditions for its
occurrence. The H-combustion flashes we observe in super-AGB stars can be
recurrent as they are initiated after the peak He-burning luminosity
of each shell flash as a result of a very fast and efficient dredge-up
while the He-shell is still convective.
Nevertheless, the nucleosynthesis processes following the ingestion
are expected to be quite similar, being driven by a fast burning of protons
and by the consequent formation of \carbon[13].
 
The \carbon[13] made by H-\carbon\ combustion activates the $\carbon[13](\alpha,n)\oxygen$ reaction, producing an intermediate neutron density of  $N_\mem{n}\approx
10^{15}\mem{cm^{-3}}$, defined as the \iprn\ by
\citet{cowan:77}. We adopt this designation for all n-capture regimes driven
by convective burning of protons in He-burning convection zones. Such
conditions can be found in numerous stellar environments, particularly
in stars of very low metallicity, such as the the He-core flash in
low-metallicity stars \citep{campbell:10}.

As shown recently by \citet{herwig:11} 1D stellar evolution
simulations with mixing-length theory for convection do not predict
surface observed abundances of the post-AGB H-ingestion star Sakurai's
object. Instead, alternate mixing assumptions had to be made for that
reactive-convective event, based on its hydrodynamic nature. Recent
attempts of hydrodynamic simulations of the H-ingestion event in
Sakurai's object have shown a much more violent behaviour than
expected from stellar evolution models, and led to the discovery of
the global oscillation of shell-H ingestion \citep[GOSH;][]{Herwig2014}. The GOSH is a
non-radial instability that develops out of a spherically symmetric
initial state, and rearranges the internal stratification in ways that
would have been impossible to predict based on stellar evolution
simulations alone. The GOSH simulations of \citet{Herwig2014} only
cover a first non-radial instability which is likely to be followed by
further violent and non-radial hydrodynamic events. Validated
sub-grid models for reactive-convective regions to be employed in
stellar evolution models are therefore not yet available, and thus we
have to consider stellar evolution results as uncertain once the
convective-reactive phases are occurring. 

With these caveats in mind we will briefly present the physics
assumptions for our simulations in Section~\Sectff{method}. In
Section~\Sectff{results} we describe the main results. In
Sections \ref{sec:consequences} and \Sectff{conclusions} we discuss the potential
consequences of our results.

\section{Method}
\label{sec:method}

\subsection{Stellar evolution code and physics assumptions}
For the stellar evolution simulations presented here we use the MESA
code \citep{paxton:11,MESA2013,MESA2015}\footnote{details about the
versions of the MESA code that were used will be made available online
along with the inlists.}.
The MESA papers by Paxton et al. already contain verification cases
for a wide range of stellar evolution scenarios, including thermal pulses
and dredge-up in AGB stars. In addition, we have now also run massive
AGB models with the same initial abundances and similar enough physics
assumptions as those chosen for the grid of models including massive
AGB stars by \citet{herwig:04a} obtained with the EVOL stellar
evolution code, and we find the MESA results again to be in good
agreement.

We use a custom nuclear network with 31 species---based on the
MESA network \texttt{agb.net}---including the major
isotopes for elements from H to Al and closing with \silicon[28].
For reactions we included PP-chains,
CNO cycles, the NeNa  and MgAl cycles, 3$\alpha\rightarrow\carbon$,
$\carbon[12](\alpha,\gamma)\oxygen[16]$, $\carbon[13](\alpha,n)\oxygen[16]$,
and $\carbon + \carbon$ (both $\alpha$ and $p$ exit channels),
amongst other less critical reactions.
Although we included the \nitrogen[13] isotope explicitly we did
not include the $\nitrogen[13](p,\gamma)\oxygen[14]$ reaction,
which would have been relevant. However, this omission will not change
any of the main findings, and will be corrected along with several
other approximations to the network in a forthcoming more detailed
nucleosynthesis investigation of these models.
Reaction rates are taken from the
NACRE compilation \citep{angulo:99}.

The mixing length theory of convection (MLT) is employed
to calculate the temperature gradient in the convective regions and predict
the convective velocities.
Mixing in convective regions is approximated
as a diffusive process with diffusion coefficient
$D=\frac{1}{3}v\alpha_\mathrm{MLT} H_P$, where
$v$ is the convective velocity according to MLT and $H_P$
the scale height of pressure.
For the mixing-length parameter we adopt
$\alpha_\mathrm{MLT}=1.73$.
In addition to the default mesh controls
we refine the mesh on the abundances of H, \helium, \carbon[13] and
\nitrogen. Additional criteria are added to resolve He-shell flashes and
the advance in time of the thin H-burning shell during the interpulse phase.
We use the atmosphere option \texttt{'simple\_photosphere'}, which gives the surface
pressure as
\begin{equation}
P_s = \frac{2GM}{3\kappa_s R^2}\left[1 +
1.6\times10^{-4}\kappa_s\left(\frac{L/L_\odot}{M/M_\odot}\right)\right]
\end{equation}
(see \citealp{paxton:11} and \citealp{CoxGiuli1968}, Section 20.1).
$\kappa_s$ is the surface opacity, for which an initial guess is made and converged
iteratively with the pressure.
We assume a mass loss rate of $\log(\dot{M} / \msun\,{\rm yr}^{-1}) \approx -4.8$
which is obtained by using the the \citet{bloecker:95a} mass loss rate in MESA with
$\eta=5\times10^{-4}$. This is likely to be at the low end of a
rather uncertain range of possible mass loss rates for these stars.
We adopt the OPAL C- and
O-enhanced opacities in MESA \citep{iglesias:96}.

\subsection{Convective boundary mixing}
\label{sec:CBM}
We have discussed convective boundary mixing (CBM) and some evidence from
hydrodynamic simulations in the Introduction. Note that we are
avoiding the term overshooting and instead prefer to describe a wide
range of hydrodynamic instabilities that can be observed at and near
convective boundaries as CBM. A variety
of fluid dynamics processes and hydrodynamic instabilities can
contribute to CBM, such as buoyancy-driven overshooting of coherent
convective systems \citep{freytag:96}, Kelvin-Helmholtz instabilities
\citep{Meakin:2007dj,casanova:11} which in turn may be enabled by boundary-layer
separation \citep{Woodward2015}, or mixing due to internal gravity
waves \citep{Denissenkov:2003gx}. 

As we described in the Introduction, our stellar evolution simulations
approximate abundance mixing due to these CBM processes---no matter what
their physical origin
is---via the exponentially decaying mixing algorithm described in
\citet{freytag:96} and first applied in stellar evolution calculations
by \citet{herwig:97}.
In this approximation, the diffusion coefficient takes the form
\begin{equation}
D_\mathrm{CBM} = D_0\exp\left(-\frac{2|r-r_0|}{fH_P}\right),
\label{eq:D-definition}
\end{equation}
where $H_P$ is the pressure scale height and $D_0$ is the diffusion coefficient
given by the mixing length theory of convection at a 
location $r_0$.
$f$ is a parameter defining the e-folding length $fH_P$ of the diffusion
coefficient.

$D_0$ is not defined at the convective boundary because in MLT, the
convective velocity is zero there. Instead of taking $D_0$ at the last convective
zone, the MESA implementation allows to specify how far inside the
convection zone from the boundary the exponential decay should begin.
The Eulerian coordinate of this location is then $r_0$.
Indeed, the results of \citet[][their Figs. 10 and 14]{Freytag1996} and
\citet[][their Fig.~24]{Herwig2006} show that the decay
of the radial component of the convective velocity and of the inferred radial
diffusion coefficient begins inside the convective zone and not at the
formal convective boundary.
In the MESA implementation, there is a second parameter for each convective
boundary $f_0$, which specifies how far from the convective boundary, on
the convective side, the decay should begin. This distance is $f_0H_P$.
In our models, we use $f_0=f$ at all convective boundaries.
The diffusion coefficient as formulated in Eq.~\ref{eq:D-definition} is
then applied in the direction of the convective boundary and across into
the stable layer.

The value of $f$ has no direct, simple and intuitive relation to
the physical flow properties---other than larger values indicate more efficient
and more spatially extended mixing than smaller values---because
it is a number that, when multiplied by the local scale height of pressure,
gives the e-folding length of the diffusion coefficient that is representing
mostly non-diffusive mixing processes caused by a variety of hydrodynamic
processes.

For our benchmark model we have assumed $f=0.014$ at all convective boundaries
(including the H- and He-core burning phases), except at the
bottom of the convective envelope (including during the third dredge-up, TDU,
if it occurs) where we assume $\fenv = 0.0035$, and at the bottom of the
pulse-driven convection zone (PDCZ) where we assume $\fPDCZ=0.002$.
The values of \fenv\ and \fPDCZ\ are lower by a factor of about 36
and 4 respectively compared to what we usually assume in our low-mass AGB
simulations \citep[which generally agree with observations in post-AGB
H-deficient stars and planetary nebulae; e.g.][]{werner:06,pignatari:13b,
Delgado-Inglada2015} and compared to
what has been shown to reproduce observational characteristics of nova
shell flashes \citep{Denissenkov2013}.
As discussed previously by \citet{herwig:03c} and \citet{goriely:04},
dredge-up may be \emph{hot} in (super-) AGB stars of low metallicity and larger
initial mass, where the H-burning at the bottom of the convective
envelope can become important already during the dredge-up
phase.
Another way to look at it is that the familiar hot-bottom
burning already starts during the third dredge-up in AGB stars of low
metallicity and larger initial mass.
As discussed briefly by
\citet{herwig:03c}, a larger CBM efficiency during the
dredge-up phase in massive AGB or super-AGB stars may result in the
formation of a corrosive H-burning flame that may terminate the AGB
through enhanced mass loss.

When hydrogen is burning at the bottom of the convective envelope,
the boundary will be stiffer (lower CBM efficiency) because of the entropy
barrier produced in the vicinity of the burning shell. Unfortunately,
without detailed hydrodynamics simulations or fluid experiments, it is
difficult to know how much stiffer the boundary becomes. In addition,
translating such a stiffness into a CBM parameter \fenv~is not straightforward.
The choice of $\fPDCZ=0.008$ below shell-flash convection zones on top of degenerate
cores is a good starting point, with a similar value being able to reproduce
the intershell abundances of post-AGB stars and the ejecta enrichment and fast
light curve rise times of novae \citep{werner:06,Denissenkov2013}.
However, the conditions in the shell flashes of super-AGB stars, post-AGB
stars and novae are certainly not identical. For example, in post-AGB stars
the peak luminosities are higher than those in super-AGB stars by roughly
an order of magnitude. Furthermore, the stiffness of the convective boundaries
and the aspect ratio of the convection will naturally also vary from site to
site.
The correlation of CBM efficiency with nuclear energy generation rate or
driving luminosity of the convection is no easy task and requires detailed,
targeted study.
For these reasons we study a range of values for \fenv~and \fPDCZ~in the
present work with a view to motivating such targeted studies.

Again, the
details will depend critically on the assumed mixing properties at
the convective boundaries which need to be investigated eventually by
means of three-dimensional stellar hydrodynamics simulations. We
address this uncertainty by providing a parameter study of the
dependence of our H-ingestion results on the CBM parameters.
As we will describe in Section~\ref{sec:results}, H-ingestion in
our TP-SAGB models is highly sensitive to CBM at the base
of the PDCZ and at the base of the hot-bottom-burning H-envelope during TDU.
These affect the response time or rate of the dredge-up as well as its depth
\citep[see][]{Mowlavi1999,Herwig2004}. The CBM \emph{above} the PDCZ, which is a
critical factor in the H-ingestion He-shell flashes of low-metallicity AGB
stars and VLTPs in post-AGB stars, is of less importance here.
Indeed, we have computed models with a range of CBM efficiencies above
the PDCZ and it is clear that this CBM does not initiate a H-ingestion
TP in our TP-SAGB models unless unjustifiably high efficiencies are used.
The overshoot at the top of the PDCZ has been studied by \citet{herwig:00c},
who also find no significant impact of this CBM on the proximity of the
PDCZ to the convective H envelope.
Thus, we limit our parameter study to only \fenv\ and \fPDCZ.

\section{Stellar evolution models}
\label{sec:results}

The evolution up to the beginning of the thermal pulse phase proceeds
in a similar manner to that described in previous studies
\citep[e.g.][]{ritossa:99,siess:07,doherty:10,ventura:11,Jones2013}.
The abundance signature of the early AGB
during the second dredge-up is a significant enhancement of He to a
mass fraction of about 0.37. This is followed in the more massive
super-AGB models by a `dredge-out' after the carbon shell burning
episodes, which can be considered as another
dredge-up event following an initial convective He-shell burning
episode that precedes the regular thermal pulses. This mixing
event brings large amounts of C and O to the surface
\citep[see, e.g.,][ and Section~\ref{sec:dredge-out}]{Ritossa1999,siess:07,
Herwig2012mltalf}.

Afterwards, the bottom of the convective envelope continues to
corrosively penetrate further into the He-burning ashes and the H-free
core is reduced by an additional $\Delta m_\mathrm{r}\approx
10^{-2}\msun$. This last dredge-up phase before the onset of regular
thermal pulses is facilitated by CBM. As described in
\citet{herwig:03c}, H is mixed into a very thin layer below the
Schwarzschild convective boundary and generates there H-burning
luminosities of the order $\log L_{\rm H}/L_\odot \approx 5.5$. This
mixing-induced H-flame will continue until the conditions for dredge-up
are removed, i.e.\ the excess energy from previous nuclear flashes and
core contraction has escaped the envelope. However, in view of
preceding and following mixing and burning events, this short phase of
H-flame core penetration does not have a large effect, except that the
transformation of C and O into \nitrogen\ is starting during this phase,
but only getting fully underway during the following thermal pulse
phase.

\subsection{H-$^{12}$C combustion during dredge-out}
\label{sec:dredge-out}
\begin{figure}
  \centering
  \includegraphics[width=1.\linewidth,clip=True,trim= 0mm 0mm 0mm
  0mm]{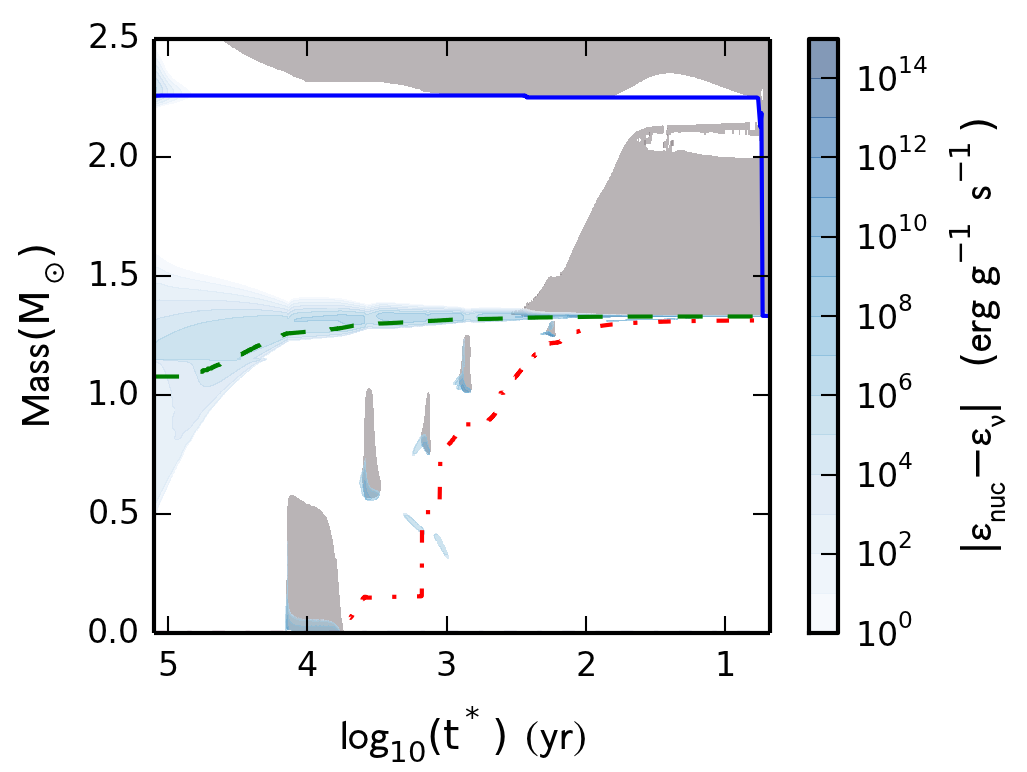}\\
  \includegraphics[width=1.\linewidth,clip=True,trim= 0mm 0mm 0mm
  0mm]{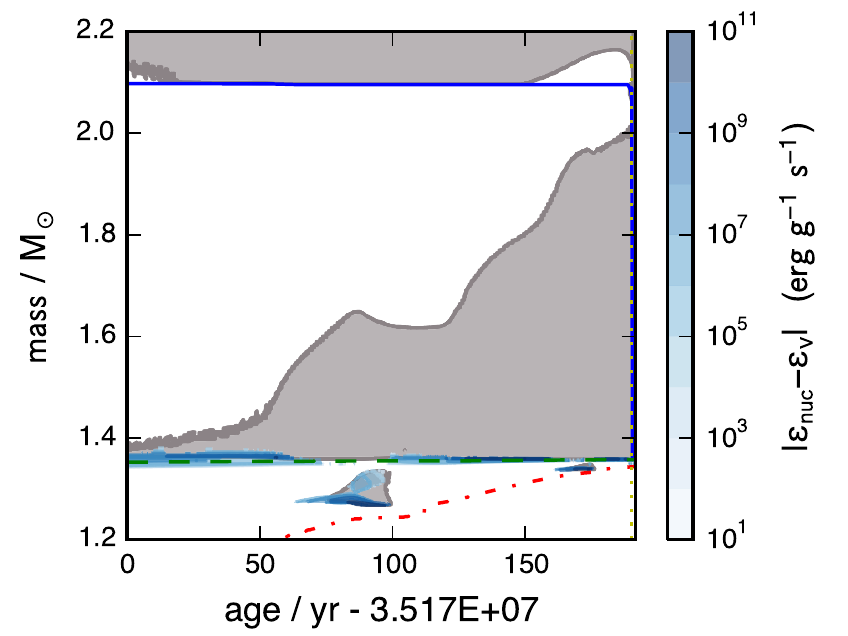}
  \caption{Top panel: evolution of the convective structure and energy generation
  during dredge-out in a 7.5\msun\ Z=0.001 stellar model. $t^*$ is the time
  remaining until the end of the calculation. Grey regions
  are convectively unstable according to the Schwarzschild criterion
  while white regions are stable. The solid (blue), dashed (green) and
  dot-dashed (red) lines indicate the H-free, He-free and
  C-free core boundaries, respectively. A convective layer develops
  in the He shell and grows outwards in mass towards the base of the
  descending convective envelope. Eventually, the two coalesce and
  protons are mixed down to the bottom of the He-burning shell
  on a convective turnover time scale. Bottom panel: same as top panel
  but for the 8.4\msun\ $Z=0.01$ model as a function of time, zoomed in.}
  \label{dredge-out-kip01}
\end{figure}

\begin{figure}
  \centering
  \includegraphics[width=1.\linewidth,clip=True,trim= 7mm 0mm 7mm
  12mm]{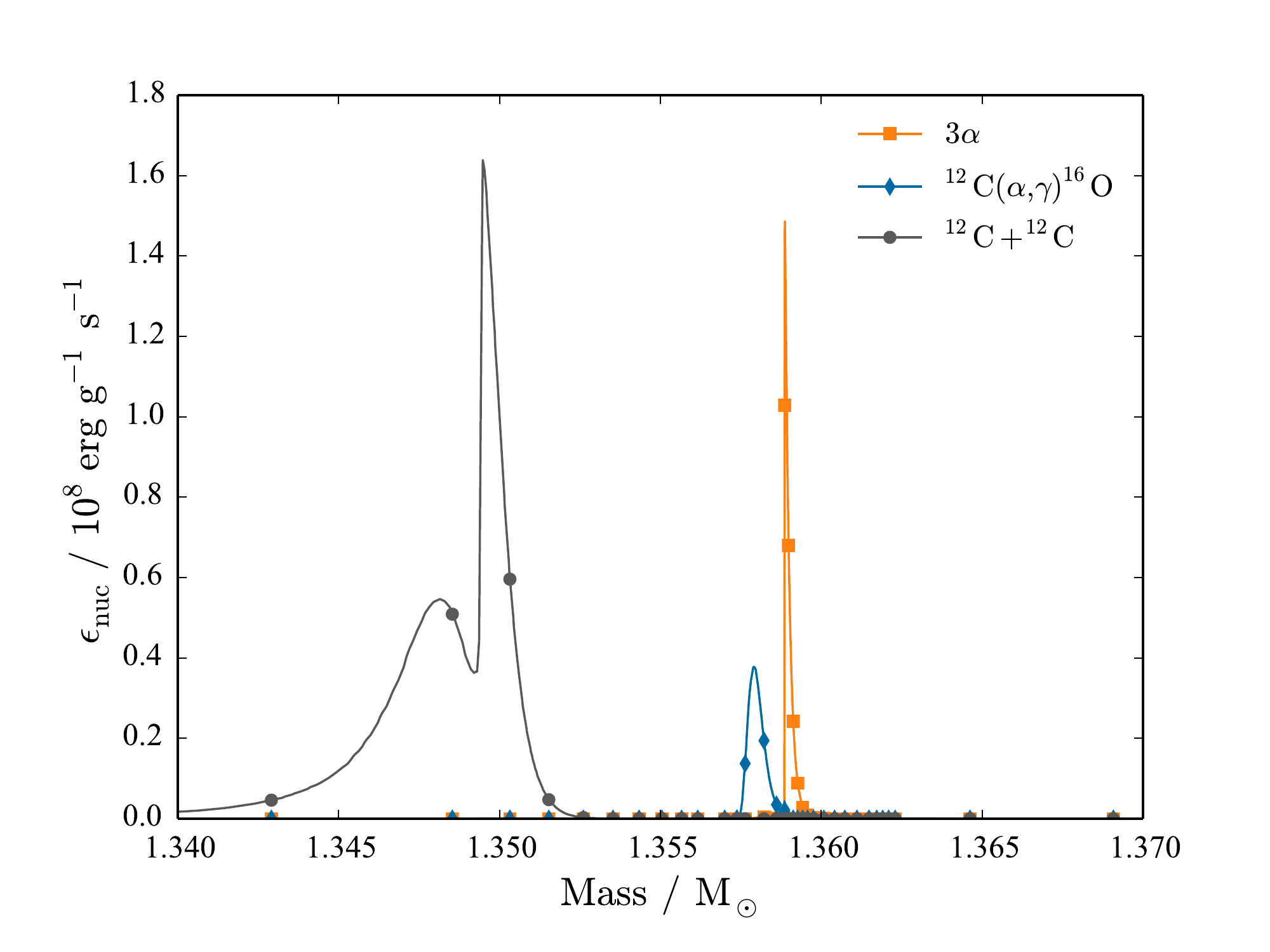}
  \caption{Stratification of the three predominant energy-generating
  nuclear reactions during the dredge-out phase in the 8.4\msun\ $Z=0.01$
  model from Fig.~\ref{dredge-out-kip01} (bottom panel) at a time from the
  right hand side of the plot where the yellow dotted line is drawn.}
  \label{energy_during_DO}
\end{figure}

The most massive super-AGB stars and failed massive stars \citep{Jones2013}
experience a dredge-out event \citep{ritossa:99,siess:07,Doherty2015}.
This is when shell
helium burning releases enough energy to induce convection during the
second DUP, as mentioned briefly above.
A combination of the release of gravo-thermal energy and
of nuclear energy cause the helium-shell convection zone to grow (outwards)
in mass and merge with the descending base of the convective hydrogen
envelope (Fig.~\ref{dredge-out-kip01}). There are three main reactions contributing to the nuclear energy
generation during this phase: $\carbon+\carbon$, $\carbon(\alpha,\gamma)\oxygen$
and the triple-$\alpha$ reaction. The energy sources are physically stratified in only a very thin
(roughly $3\times10^{-2}\msun$) mass shell (see Fig.~\ref{energy_during_DO}).

The two main consequences of dredge-out are the enrichment of the envelope
with  ashes of hydrogen burning and incomplete
He burning, and the advection of protons down to He-burning
temperatures where primary \carbon\ is present with a typical mass
fraction of $X(\carbon)\approx0.5$. A strong flash occurs as the protons
are mixed rapidly into the hotter layers below the base of the H
envelope. This has previously been reported by \citet{Gil-Pons2010},
who found that the resulting hydrogen burning produced local
luminosities in excess of $10^6~\Lsun$. As we will describe here our models suggest that in rapid combustion of the H with \carbon\ local luminosities well in excess of $10^9~\Lsun$ may be reached.

The dredge-out event is illustrated in Fig.\,\ref{dredge-out-kip01} for a
7.5\msun\ $Z=10^{-3}$ model (top panel) and for a 8.4~\msun\ $Z=0.01$ model
(bottom panel).
In this simulation the strict Schwarzschild
criterion (i.e.\ no CBM) is
assumed at the lower boundary of the convective envelope.
Our simulations confirm that dredge-out is encountered without CBM at
the lower boundary of the convective envelope.
Dredge-out has been reported and described also by, e.g., \citet{ritossa:99},
\citet{siess:07}, \citet{Gil-Pons2010} and \citet{Doherty2015},
none of whom prescribe for CBM (though each group has their own method of
treating the convective boundary).
We have simulated massive super-AGB and failed massive stars with
initial metallicities between $Z=10^{-5}$ and 0.02, and dredge-out is
seen to occur at all metallicities when holding our CBM assumptions
described in Section~\ref{sec:CBM} \emph{and} when setting $\fenv=0$.

\begin{figure*}
  \centering
  \includegraphics[width=1.\linewidth,clip=True,trim= 7mm 0mm 7mm
  10mm]{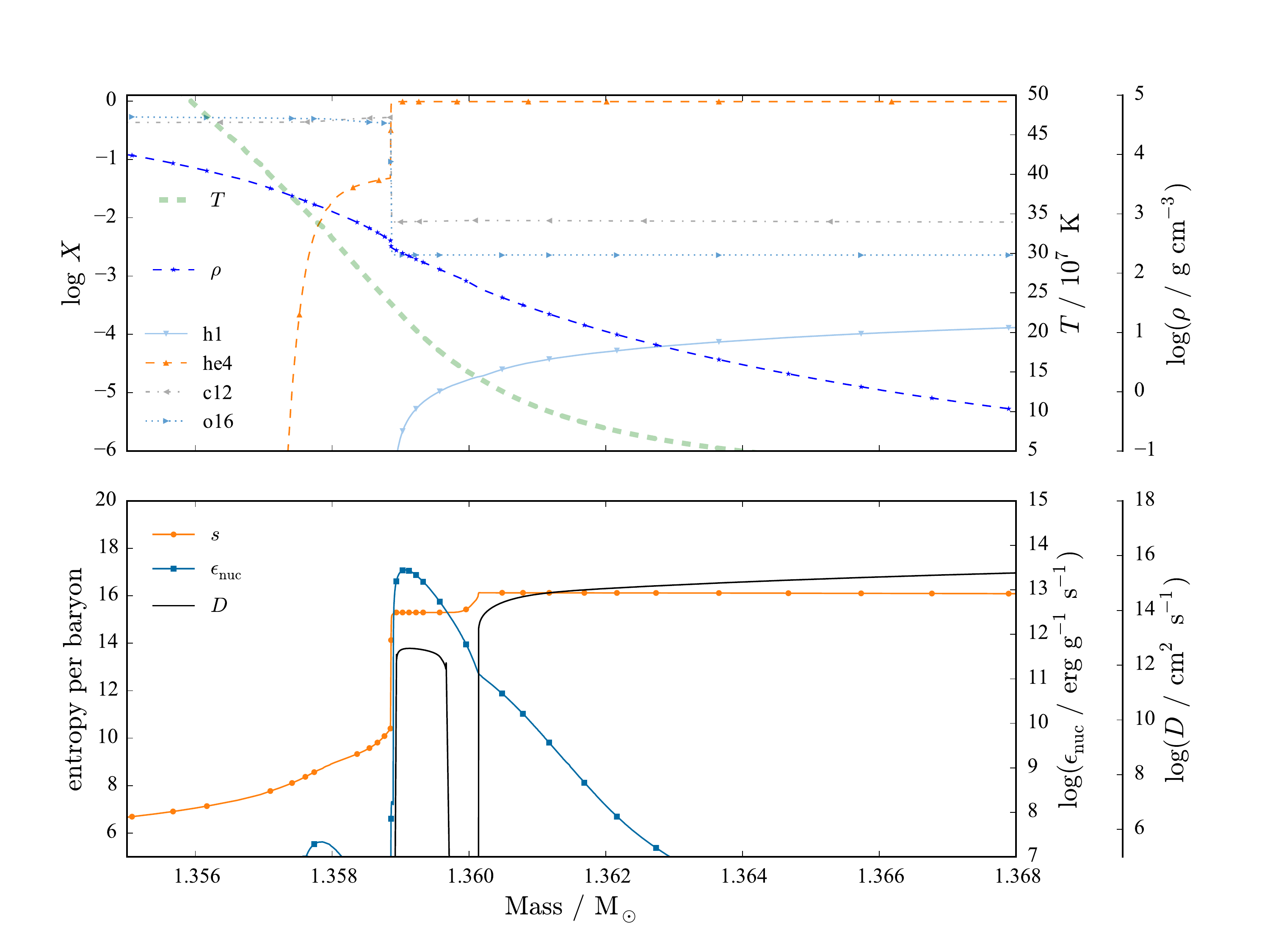}
  \caption{Abundance profiles with the temperature and density stratification
  (top panel), entropy $s$, specific energy generation rate and
  and diffusion coefficient (bottom panel) in the 8.4\msun\ $Z=0.01$
  model during the peak energy generation due to H-$^{12}$C combustion
  (very right hand side of Fig.~\ref{dredge-out-kip01}, bottom panel,
  where the solid blue H-free core boundary penetrates down to the green
  dashed He-free core boundary, indicated by a vertical dotted yellow line).
  The points are illustrative and do
  not reflect the grid.
  At the base of the convective helium
  layer (just below mass coordinate 1.360\msun), the mass fraction of
  protons is a few $10^{-6}$ and the temperature is in excess of $2\times10^8$~K.
  These are precisely the conditions under which the neutron densities 
  characteristic of the $i$-process can be achieved due to the release
  of neutrons in the $\carbon[12](p,\gamma)\nitrogen[13](\beta^+\nu)\carbon[13](\alpha,n)\oxygen[16]$
  reaction chain. In our 1D hydrostatic simulations the
  protons burn as they are transported inwards through regions of increasing temperature. The
  burning taking place at mass coordinate $M_r\approx1.360\msun$ increases the entropy
  and causes a split in the convection zone, which can be seen as a break in
  the diffusion coefficient profile.}
  \label{dredge-out-prof01}
\end{figure*}

During the dredge-out, protons are
mixed down through the helium shell toward higher temperatures.
In the $8.4\msun$,  $Z=0.01$ model the protons reach temperatures
in excess of $\natlog{2}{8}\mathrm{K}$, where their fraction by mass is a few $10^{-6}$. The
energy generation at this location reaches a few $10^{13}$ erg g$^{-1}$ s$^{-1}$.
The abundance profiles, temperature stratification and specific nuclear energy
generation at the time of peak nuclear energy generation are shown in
Fig.\,\ref{dredge-out-prof01} and these characteristics are summarised again
in Table~\ref{tab:summary}.

\begin{table}
\centering
\def\arraystretch{1.5}
\begin{tabular}{l c c c c}
\hline
site & $T\,/\,\mathrm{K}$ & $\rho\,/\,\mathrm{g\,cm}^{-3}$ & $X_\mathrm{H}$ & $X_\mathrm{C}$ \\
\hline
dredge-out         & $\sim 2\times10^8$   & $\sim 2\times10^2$ & $\gtrsim 10^{-6}$ & $\sim10^{-2}$ \\
TP-SAGB         & $\sim 2\times10^8$ & $\sim60$ & $\sim 3\times10^{-2}$ & $\sim0.2$ \\
\hline
\hline
\end{tabular}
\caption{Characteristic conditions during H-ingestion in both dredge-out and TP-SAGB sites.}
\label{tab:summary}
\end{table}

In order to characterize the dynamic importance of the convective-reactive
nuclear energy
release we introduce a dimensionless number which relates the nuclear
energy released to the internal energy. As protons are advected into
the He-shell convection they will encounter increasing temperature
and primary \carbon\ with mass fraction well above $10\%$. A large
amount of nuclear energy is released on the convective advection time
scale via the $\carbon(\proton,\gamma)\nitrogen[13]$
reaction \citep[see appendix A.2 in][]{herwig:11}. The fastest time scale
on which this energy can escape under the assumption of hydrostatic equilibrium is
the convective turn-over time. Thus, the product $\epsilon_{\rm
  nuc}\tau_{\rm conv}$ is here an estimate of the nuclear energy from
convective-reactive H combustion that will accumlate in the
combustion flame. We define
\begin{equation}
\label{eq:h-number}
H = \frac{\epsilon_{\rm nuc}\tau_{\rm conv}}{E_{\rm int}},
\end{equation}
where $\epsilon_{\rm nuc}$ is the specific energy generation from nuclear
reactions, $\tau_{\rm conv}$ is a mixing timescale and $E_{\rm int}$ is the
specific internal energy of the stellar material. For the mixing time
scale, using the scale height of pressure at the base of the convective
He-burning shell ($H_P \approx  1000\mathrm{km}$) as a length scale
and the \emph{maximum} convective velocity predicted by MLT inside the
helium-burning shell ($v_\mathrm{conv} \approx 3 \mathrm{km s^{-1}}$),
we find a lower limit of
\begin{equation}
\tau_{\rm conv} \approx \frac{H_P}{v_\mathrm{conv}} = 409\mathrm{s}.
\end{equation}
With these asumptions $H = 0.11$. This means that in the combustion
flame layer the cumulative nuclear energy input is
of the order of $ 11\%$ of the internal energy, or 8\% of the binding
energy of the layer, or greater. 

The $H$-number estimate suggests that the H-ingestion may release and add
to the flow a significant fraction of the binding energy of the layer,
in which case fundamental assumptions of stellar evolution, such as
hydrostatic equilibrium and the applicability of MLT become
inappropriate. \citet{herwig:11} have shown that in another case of
convective-reactive H-ingestion observations can not be reproduced
with MLT/stellar evolution models. Sakurai's object is a
very-late thermal pulse object with a $2$-$\mathrm{dex}$ enhancement of
first-peak n-capture elements. Only by adopting significant
modifications to the MLT mixing predictions could nucleosynthesis simulations
reproduce the observations. \citet{Herwig2014} presented 
hydrodynamic simulations of the H-ingestion event in Sakurai's object
and reported a non-radial global oscillation of shell H-ingestion
(GOSH) for a situation similar to the one encountered here for the
H-ingestion during the dredge-out. We therefore conclude in a similar manner
that these stellar evolution models with MLT, and estimates such
as as the $H$-number based on these models, are not entirely
realistic once the H-ingestion event with a large $H$-number occurs.

The peak energy generation due to H-$^{12}$C combustion during the dredge-out is short-lived
in our MLT stellar evolution models. The burning of
protons as they are transported inwards modifies the entropy structure of the region of interest.
The entropy and diffusion coefficient profiles are shown in Fig.\,\ref{dredge-out-prof01}.
In the stellar evolution model the entropy produced by the burning of
hydrogen causes a split
in the convection zone, resulting in two convective
regions that do not mix further. The protons in the lower convective zone are quickly destroyed,
after which fresh hydrogen brought down is burned at the top of the split. There, the temperature
is only $\approx \natlog{1.5}{8}~\mathrm{K}$ but nevertheless a few
$10^{-3}$ in mass fraction of $^{13}$C is produced. 
A split in the diffusion coefficient under similar conditions was seen
in the models of \citet{herwig:11}. The authors found that the
basic assumptions that contribute to the formation of the split---such
as a spherically symmetric H-abundance and a spherically symmetric inward
diffusion of H---were not found in three-dimensional 
hydrodynamic simulations. Hence the hypothesis that two
convective regions should mix to some extent. 

This modified mixing assumption by \citet{herwig:11}
(whereby the two convection zones
split by the formation of the entropy barrier due to H-burning are allowed to 
continue to mix) leads to n-capture nucleosynthesis at neutron
densities of $N_\mathrm{n} \approx \power{10}{15}\mathrm{cm^{-3}}$.
This is approximately two orders of magnitude higher than in the highest
neutron densities encountered in the \spr\ powered by the \neon[22]
neutron source in the He-shell flash convection zone of massive AGB
stars \citep[see, e.g.,][their Table~1]{Straniero2014}.
This n-capture regime has been labelled the intermediate regime,
or \ipr\ by \citet{cowan:77}. We propose that the H-ingestion induced hydrodynamics in
the super-AGB dredge-out He-burning convection zone would behave
similarly to the situation investigated for the H-ingestion event in
Sakurai's object, in the sense that further mixing between the
convecion zones will lead in a similar way to the copious production of
\carbon[13] and the associated intense neutron burst (see, e.g.,
\citealp{Lugaro2009,campbell:10}). This hypothesis
will need to be tested by appropriate hydrodynamic simulations. In the
meantime, based on the evidence available at this time we adopt
as the most likely scenario that the dredge-out in super-AGB stars
is a site for \ipr\ nucleosynthesis (see \Sect{consequences}).

In failed massive stars that would become electron-capture supernovae, the dredge-out occurs
about 30 years before the explosion. With hydrodynamics simulations suggesting that the flow
of protons into the He-burning layer should persist
it is entirely possible that enough energy is released in the ingestion to unbind and
eject material that has been exposed to such large energy generation rates.
Without detailed multi-physics simulations it is very difficult to know to what extent
energy is dissipated during the ingestion/ejection event, but nevertheless
we entertain this possibility and examine it in more detail in \Sect{mass_ejection},
where we also predict what the light curve of such a transient might look like.

\subsection{TP-SAGB models and thermal pulse evolution}
\label{sec:tp-evol}
We have calculated TP-SAGB models with initial metallicities of $Z=0.02$, 0.01,
$10^{-3}$, $10^{-4}$ and $10^{-5}$ (Table~\ref{tab:5model_table}). These models
were computed holding the input physics assumptions described in
Section~\ref{sec:CBM}.
The thermal pulse properties are sensitive to the core mass, and
therefore the initial masses were chosen such that the CO core
masses of all the models at the first thermal pulse were similar
(the standard deviation in the CO core masses at the first thermal pulse
is 0.012\msun; see Table~\ref{tab:5model_table}).

\begin{figure*}
\centering
\includegraphics[width=\linewidth]{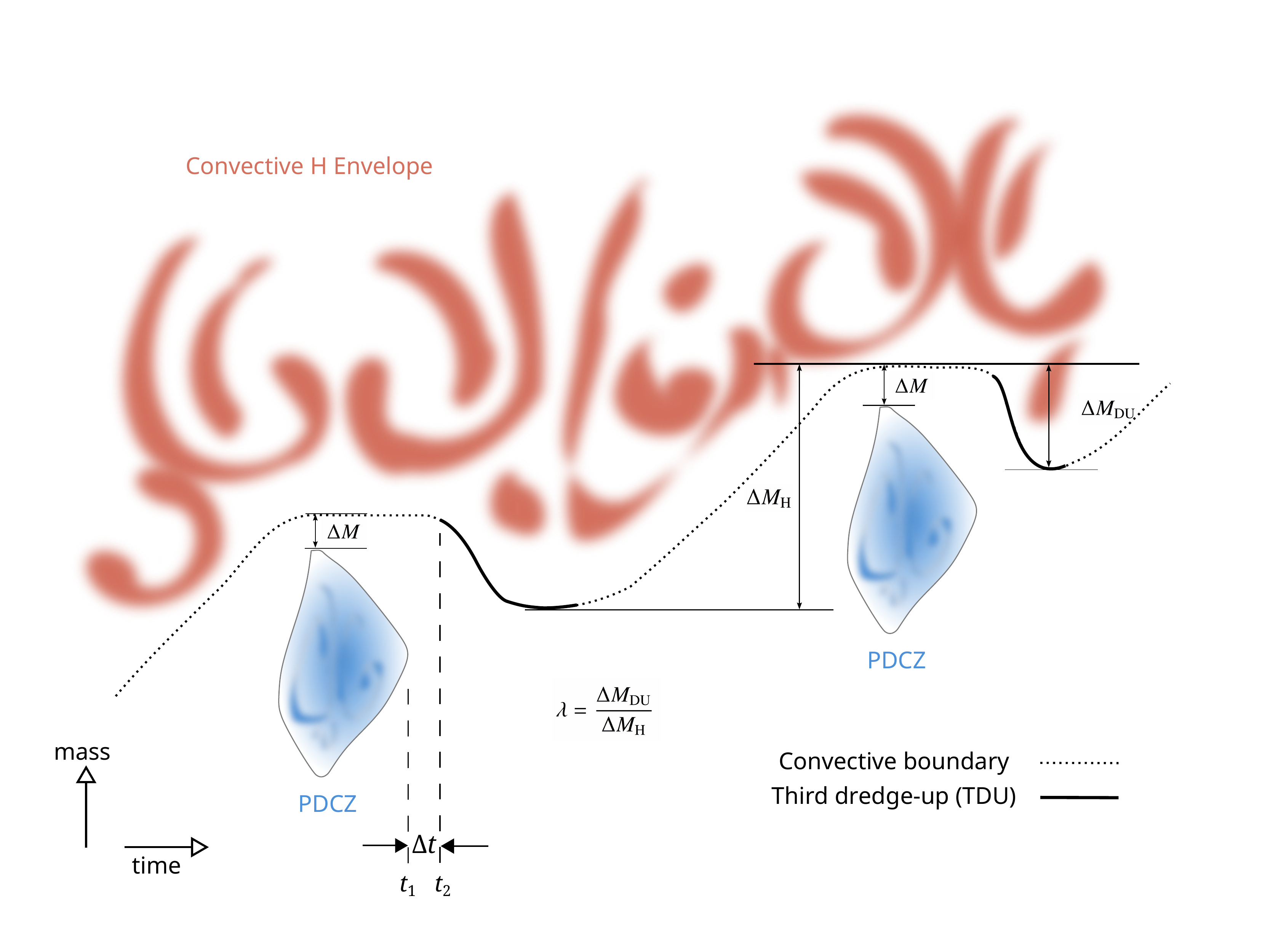}
\caption{Illustration of the TP-SAGB phase, defining the pulse-driven
  convection zone (PDCZ), the convective envelope base penetration (CEBP),
  $t_1$ and $t_2$ -- the times of PDCZ extinction and beginning of CEBP,
  respectively -- and the time interval between $t_1$ and $t_2$, $\Delta t$.}
\label{fig:TP_definitions}
\end{figure*}

The purpose of this metallicity survey in the models is to compare the
thermal pulse properties of the super-AGB models at different metallicities with
regards to approaching the conditions for a H-ingestion event to take place.
H-ingestion occurs in our TP-SAGB models, as we will show, when the
third dredge-up (TDU; the deepening of the convective envelope in mass
following the thermal pulse; see Fig.~\ref{fig:TP_definitions}), facilitated by
CBM, reaches into the pulse-driven convection zone (PDCZ). In order for this\
to happen, the TDU must be rapid enough, and deep enough, to cause a
convective exchange of material between the convective H-envelope and the
PDCZ. 

\begin{table}
\def\arraystretch{1.5}
\centering
\begin{tabular}{c c c c c}
\hline
$Z$ & $M_\mathrm{ini}/\msun$ & $M_\mathrm{C}^\mathrm{SAGB}$ & H ing? & 
peak $L_{\rm H}~/~10^9L_\odot$ \\
\hline
0.02         & 8.2   & 1.26975 & N & - \\
0.01         & 8.15 & 1.28025 & N & - \\
$10^{-3}$ & 7.0   & 1.24606 & Y & 8.25 \\
$10^{-4}$ & 7.0   & 1.26263 & Y & 8.26 \\
$10^{-5}$ & 7.2   & 1.27432 & Y & 8.48 \\
\hline
\hline
\end{tabular}
\caption{Summary of the models presented here as a metallicity study.
These models assume $\fenv=0.0035$ and $\fPDCZ=0.002$ (see text for details).
$M_\mathrm{C}^\mathrm{SAGB}$ is the mass of the H-free core at the
first thermal pulse. Also given is whether or not a H-ingestion TP is
encountered and if so, the peak H-burning luminosity is given.}
\label{tab:5model_table}
\end{table}

\begin{figure}
  \centering
    \includegraphics[width=1.\linewidth,clip=True,trim= 0mm 0mm 0mm
  0mm]{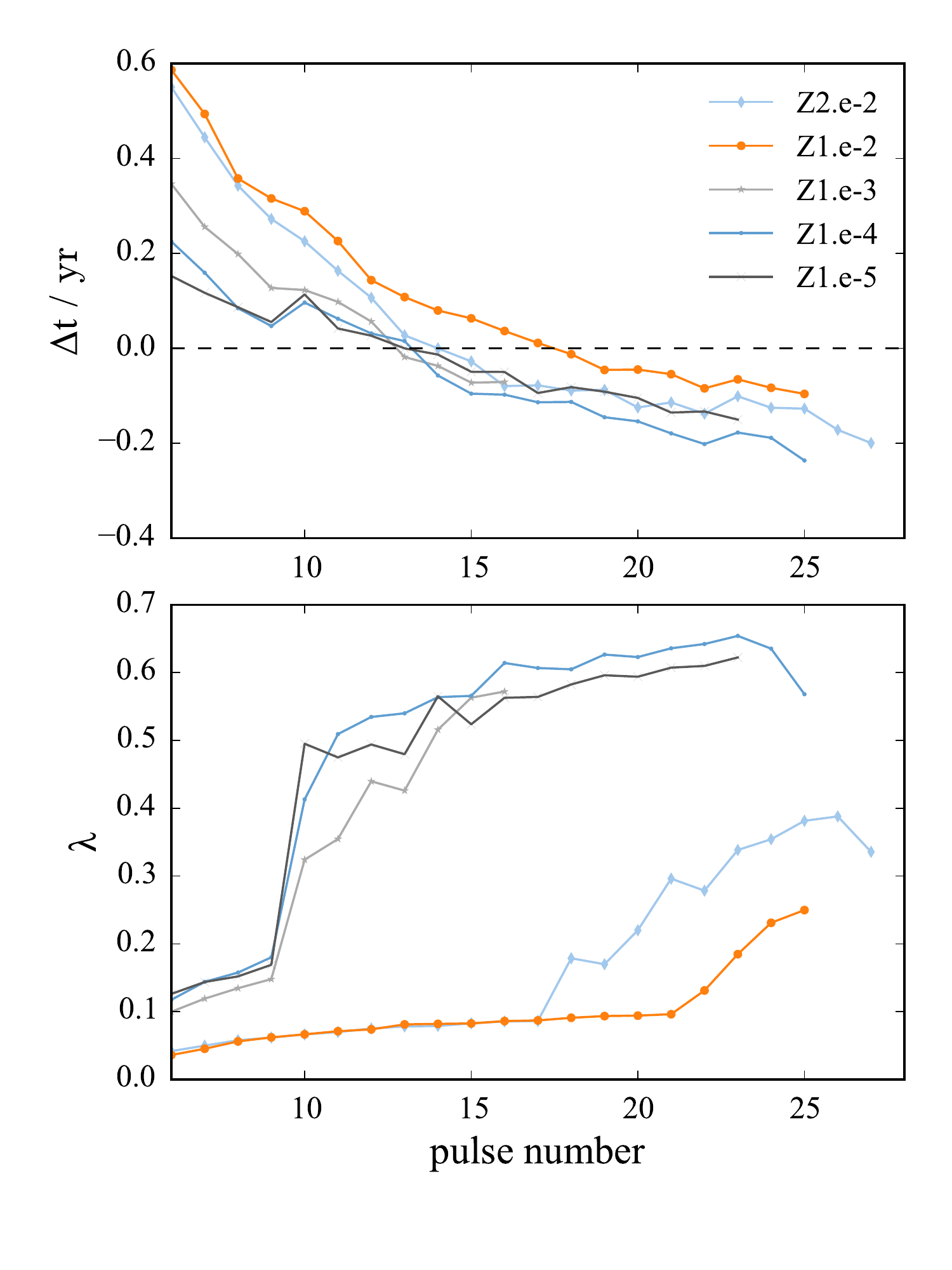}
  \caption{\textit{Top Panel:} Time between the extinction of the
  pulse-driven convection zone (PDCZ) and the beginning of third dredge-up (TDU), $\Delta t$
  (see Fig~\ref{fig:TP_definitions} and the text for details).
  Hence, above the horizontal dashed line at $\Delta t=0$ the TDU begins
  after the extinction of the PDCZ and below the
  line $\Delta t=0$ the TDU begins while the PDCZ is still active.
  \textit{Bottom panel:}
  TDU efficiency $\lambda$---the ratio of the mass of material mixed up to the
  amount the H-free core grew during the preceding H-burning inter-pulse
  phase---as a function of pulse number for different metallicity models.}
  \label{TP_DUPs_DT}
\end{figure}

A TDU event regularly follows each He-shell flash although is
often absent for the first few flashes.
As the models evolve along the TP-SAGB, the time $\Delta t$ between the
extinction of the PDCZ $t_1$ and the beginning of the TDU $t_2$ becomes shorter
(see definitions in the illustration in Fig.~\ref{fig:TP_definitions}).
This trend is seen for models at all metallicities that we have considered and
is shown in Fig.\,\ref{TP_DUPs_DT} (top panel; see also \citealp{Mowlavi1999}
and \citealp{Herwig2000}).
We define the beginning of the TDU in these
models as the time when the base of the convective envelope has deepened in mass
by 1\% of $\Delta M_\mathrm{DU}$ for a given TDU.
Of particular note is the result that after about 10--15 pulses (depending on the initial metallicity)
the TDU starts while the PDCZ
is still active, with the lower metallicity models reaching this point earlier in the TP-SAGB.
In this case $\Delta t < 0$ (Fig.\,\ref{TP_DUPs_DT}, top panel), which is
one of the criteria for hydrogen from the envelope to be ingested into the PDCZ.
If $\Delta t$ were always positive then the kind of ingestion that we
describe here, where the hot TDU reaches the PDCZ, could never happen.
Even if we do not always see explicitly the H-ingestion in 1D models,
one of the main conditions for this to happen---i.e.\ the TDU beginning already
at the time when the PDCZ is still
active---is given robustly after at most 16 thermal pulses at all initial metallicities for which we have
computed models (Table~\ref{tab:5model_table} and Fig.~\ref{TP_DUPs_DT},
top panel).

The TDU efficiency $\lambda$ is defined as the ratio of dredged-up mass $\Delta M_\mathrm{DU}$ to
core-growth $\Delta M_\mathrm{H}$ since the end of the previous TDU
\citep[see, e.g.,][and Fig.~\ref{fig:TP_definitions}]{Lugaro2003}.
In our models TDU is not always deep enough to mix He-burning products
to the surface (i.e.\ $\Delta M_\mathrm{DU}$ is not always larger than
$\Delta M$).
For $\Delta M_\mathrm{DU}<\Delta M$, neither does the model experience a
H-ingestion thermal pulse.

$\lambda$ is consistently higher for
the models with $Z \leq 10^{-3}$ than higher-$Z$ models for a similar pulse
number. In fact, the models with $Z \leq 10^{-3}$ all experience
efficient TDUs from the fifteenth thermal pulse at the latest,
with $\lambda>0.5$.
Indeed, in this set of models with $\fenv=0.0035$ and $\fPDCZ=0.002$ we
find that the
models with $Z\leq10^{-3}$ experience H-ingestion TPs
with peak H-burning luminosities reaching almost $10^{10}~\mathrm{L}_\odot$
(Table~\ref{tab:5model_table}).
The $Z=0.02$ and 0.01 models also show marked increases in
$\lambda$ further along the TP-SAGB but in our simulations up to the 26th and 25th thermal pulses, respectively, $\lambda<0.5$ and no H-ingestion events were
encountered.
An important implication of large $\lambda$ values in super-AGB
stars is the narrowing (or disappearance) of the electron-capture supernova
channel due to a suppressed core growth rate.

The CBM parameter $\fenv$ (see \Sect{CBM}) can be calibrated for the lower envelope convection boundaries during the TDU for low-mass AGB stars to reproduce the required partial mixing between the H-rich envelope and the \carbon-rich core to form a sufficiently large \carbon[13]-pocket for the \spr. \citet{Lugaro2003} have proposed that a CBM parameter $\fenv=0.128$ would reproduce \spr-observables.
However, \citet{pignatari:13b} have shown that with $\fenv = 0.126$ the galactic C-rich stars with the highest s-process enrichments are not reproduced (see e.g., \citealp{abia:02} and \citealp{Zamora2009} for observations).
The CBM at the bottom of the convective envelope during the TDU in high-mass AGB and super-AGB stars leads to
hot dredge-up (see \Sect{CBM}). The appropriate value for $\fenv$ is not
known for this situation, but numerical experiments show that adopting the
same value that was calibrated to satisfy the constraints of a sufficiently
sized \carbon[13]-pocket in low-mass stars would lead to very intense
H-burning that would lead to rapid disintegration of the AGB star
\citep{herwig:04a}. In our benchmark model we used a much smaller value
of $\fenv=0.0035$. Generally speaking, a much smaller value for hot DUP
compared to regular TDU is justified because the energy released by burning
protons mixed into the hot core will add buoyancy to the fluid elements
which in turn will reduce the efficiency of the CBM process. While the
efficiency of third dredge-up depends only very weakly on $\fenv$ in
low-mass AGB stars \citep{Mowlavi1999}
this is entirely different when the dredge-up is hot.
In this case the dredge-up efficiency is strongly dependent on $\fenv$,
as we will show in the following sections (see Fig.~\ref{fig:lambda-fenv}).
The hot DUP, particularly with CBM, is a difficult phenomenon to
simulate. Although it is approximated in our 1D code, the coupled behaviours
of the burning and mixing together with the structural readjustment of
the star following the peak He-shell flash luminosity is itself something that
warrants further in-depth investigation \citep[see also][]{Herwig2004}.

\begin{figure}
  \centering
  \includegraphics[width=\linewidth,clip=True,trim= 0mm 0mm 0mm
  0mm]{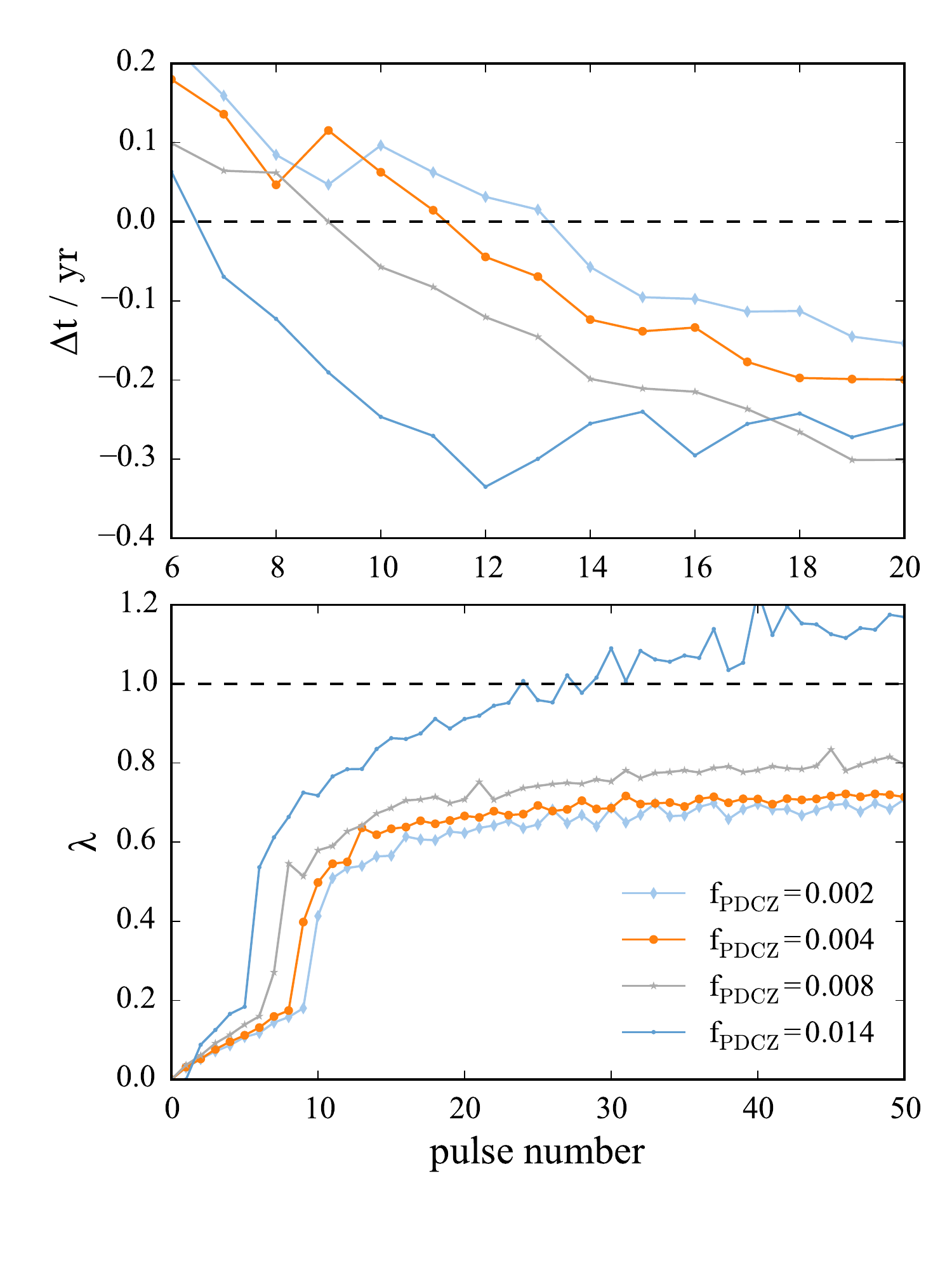}
  \caption{A series of models with $M=7~\msun$ and $Z=10^{-4}$ in which $\fenv=0.0035$ and \fPDCZ\ is increased from 0.002
  through 0.014 (first 4 entries in Table~\ref{tab:modelprops-sj}). \textit{Top panel}: same as Fig.\,\ref{TP_DUPs_DT} (top panel). \textit{Bottom panel}: same as
  Fig.\,\ref{TP_DUPs_DT} (bottom panel).}
  \label{fig:impact-fpdcz}
\end{figure}

\subsection{Impact of convective boundary mixing on the TP-SAGB phase and
hydrogen-ingestion TPs}
\label{sec:CBM_impact}
The efficiency and extent of convective boundary mixing during the TP-SAGB is not known at present.
There are little to no observational diagnostics that really help to constrain such mixing.
It is thus important to study the impact of the choice of the e-folding lengths of the
exponentially decaying diffusion coefficient ($fH_P$ in our CBM scheme) on
the occurrence and behaviour of hydrogen ingestion events during the TP-SAGB. In order to
accomplish this, we have computed a series of models from the end of the second
dredge-up in our 7\msun\ model with initial metallicity $Z=10^{-4}$. We focus our study on the
impact of two convective boundary mixing parameters: \fPDCZ---the $f$ value characterising the
mixing at the base of the pulse-driven convection zone---and \fenv---the equivalent value for the
base of the convective envelope including during its penetration following the thermal pulse (TDU).
The results of these models are
summarised in Table\,\ref{tab:modelprops-sj}. Initially, we continued the computation of the
model holding our original assumption for the convective boundary mixing scheme
($\fenv=0.0035$, $\fPDCZ=0.002$; see
Section~\ref{sec:CBM}) and found that the model indeed encountered a situation whereby
protons from the convective envelope were ingested into the PDCZ.
This He-shell flash H-ingestion is different than the equivalent event in low metallicity low- and
intermediate-mass stars \citep{fujimoto:00,herwig:03a,suda:04,campbell:08,Cristallo:2009cu},
or H-deficient post-AGB stars \citep{herwig:11}. In those situations, the hydrogen-burning
shell separating the hydrogen-rich and helium-rich material is either weak or completely
inactive. The entropy barrier is thus weaker and it is easier for the PDCZ to ingest protons.
In super-AGB stars where dredge-up is hot (i.e.~the hydrogen-burning shell is still active),
it is much more difficult for the PDCZ to simply engulf some of the hydrogen-rich material in
the same way.
Instead, the combination of effects from the CBM below the PDCZ and
the CBM below the convective envelope can be enough, in our current
parameterisations, to initiate a convective-reactive event.

\begin{figure}
  \centering
  \includegraphics[width=1.\linewidth,clip=True,trim= 8mm 0mm 0mm 
  8mm]{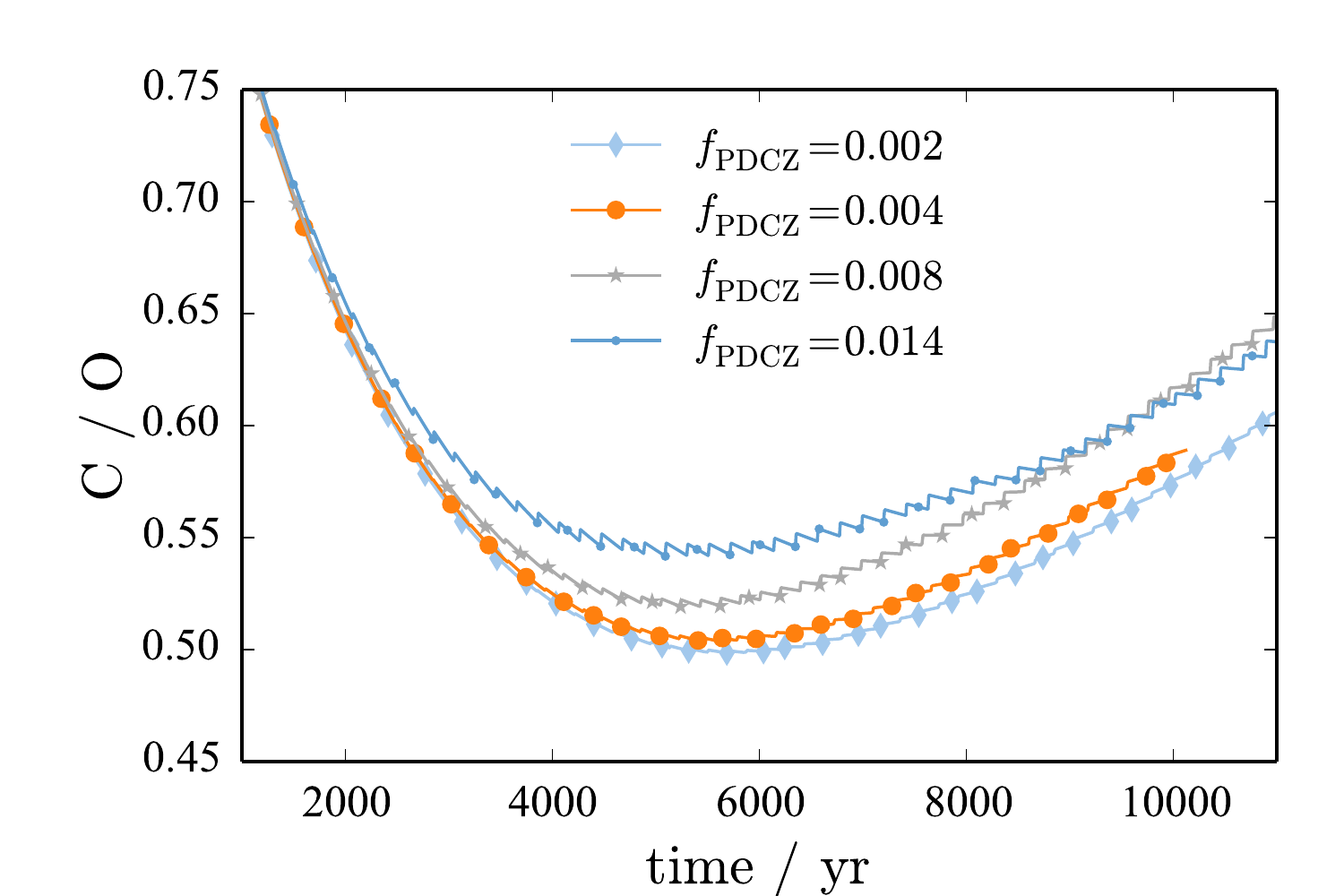}
  \caption{C/O ratio (${\rm C/O}=16X(\carbon)/12X(\oxygen)$) at the stellar surface
  for the series of models with $M=7~\msun$ and $Z=10^{-4}$ in which $\fenv=0.0035$ and \fPDCZ\ is increased
  from 0.002 through 0.014 (first 4 models in Table~\ref{tab:modelprops-sj}).}
  \label{fig:CO-fpdcz}
\end{figure}

\begin{figure}
  \centering
  \includegraphics[width=1.\linewidth,clip=True,trim= 8mm 0mm 0mm 10mm]{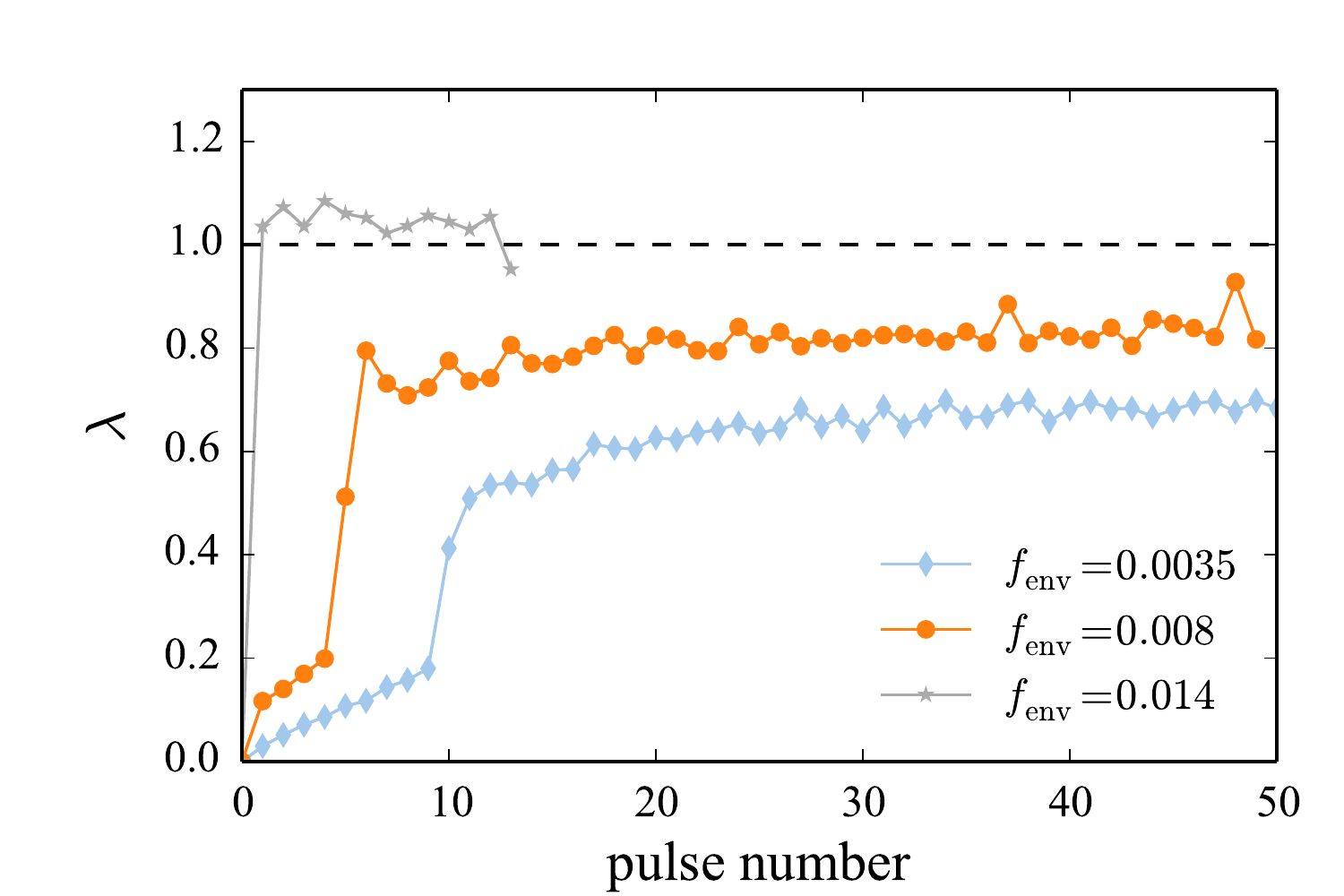}
  \caption{TDU efficiency $\lambda$ for the series of models with $M=7~\msun$ and $Z=10^{-4}$ in which
  $\fPDCZ=0.002$ and \fenv\ is increased from 0.0035 through 0.014 (first entry and middle three entries in Table~\ref{tab:modelprops-sj}). Model with $\fenv=0.022$ is actually not shown because it does not experience any TP events.}
  \label{fig:lambda-fenv}
\end{figure}

\begin{table*}
\def\arraystretch{1.5}
\centering
\begin{tabular}{c c p{7cm} c c c}
\hline
$f_\mathrm{PDCZ}$ & $\fenv$ & comments on H ingestion events& peak $L_{\rm H}~/~10^9L_\odot$ & $\lambda$\\
\hline
0.002 & 0.0035  & & 8.26 & $<1$ \\
0.004 & 0.0035  & & 3.98 & $<1$ \\
0.008 & 0.0035  & & 4.35 & $<1$ \\
0.014 & 0.0035  & PDCZ mixes up some CO core & 17.4 & $>1$ \\
\hline
0.002 & 0.008    & & 6.06 & $<1$ \\
0.002 & 0.014    & Only 14 pulses computed & 3.87 & $>1$ \\
0.002 & 0.022    & No TP. H shell corrosively burns into CO core& -- & -- \\
\hline
0.004 & 0.008    & & 7.47 & $<1$ \\
0.008 & 0.014    & Only 13 pulses computed, PDCZ mixes up some CO core & 5.84 & $>1$ \\
0.014 & 0.014    & Only 14 pulses computed, PDCZ mixes up some CO core; one pulse is shown in Fig.~\ref{fig:h-ing-kip-f-14-14} & 23.1 & $>1$ \\
0.014 & 0.022    & No TP. H shell corrosively burns into CO core & -- & -- \\
\hline\hline
\end{tabular}
\caption{Key properties of the models from the mixing parameter study at
$Z=10^{-4}$ with $M_\mathrm{ini}=7\msun$.
Aside from the models where no thermal pulses (TPs) occur, all
of the models listed here experience H-ingestion events.
$f_\mathrm{PDCZ}$ is the convective boundary
mixing parameter $f$ (Eq.~\ref{eq:D-definition}) below the pulse driven convection zone and $\fenv$ is
the parameter at the base of the convective envelope.
The peak value of the H-burning luminosity is given in units of $10^9~L_\odot$
and whether the TDU efficiency $\lambda$ is greater or less than unity is given in the final column.}
\label{tab:modelprops-sj}
\end{table*}

\subsubsection{Fixed \fenv}
Fixing the value of $f$ parameter at the base of the envelope at our
original assumption of $\fenv=0.0035$ (see above), \fPDCZ\ was increased
through the range 0.002--0.014. The
impact on the TDU efficiency $\lambda$ and the time between
the onset of TDU and extinguishing of the PDCZ is shown in the bottom panel of
Fig.\,\ref{fig:impact-fpdcz}. The peak He-burning
luminosity during the shell flash is higher for larger values of \fPDCZ. As a result the TDU
begins sooner (Fig.\,\ref{fig:impact-fpdcz}, top panel) and becomes more efficient
(Fig.\,\ref{fig:impact-fpdcz}, bottom panel).
All of the models in this series experience
H-ingestion during the TP-SAGB. Once this begins to occur, it becomes rather difficult
to define at which point the PDCZ has extinguished. It is for this reason that only up to the
20$^{\rm th}$ pulse is plotted in the top panel of Fig.\,\ref{fig:impact-fpdcz}.

The peak hydrogen-burning luminosity $L_{\rm H}$ is given for each model in
Table\,\ref{tab:modelprops-sj} (first 4 entries).
All of the models encounter at least one thermal pulse with
$L_{\rm H}>10^9~L_\odot$ and in the case with the largest extent of CBM
($\fPDCZ=0.014$) the H-burning luminosity even exceeds $10^{10}~L_\odot$.
That case is also unique because it is the only model in the series that
mixes material up from the CO (He-free) core into the PDCZ during the He
shell flash. This begins to happen at around pulse number 40. The C/O ratio
for these four models as a function of evolutionary time is shown
for the TP-SAGB phase in Fig.\,\ref{fig:CO-fpdcz}. The dredge-up efficiency
is greater than unity for only the model with
$\fPDCZ=0.014$, indicating that for such a parameterisation of the CBM
below the pulse-driven convection zone the CO core will not grow to the
critical mass for electron-captures to trigger its collapse.

\subsubsection{Fixed \fPDCZ}
A second series of models was computed in which we fixed the value of the
$f$ parameter below the PDCZ at our initial assumption of $\fPDCZ=0.002$
and increased \fenv\ through the range 0.0035--0.022. For the model with the
largest value of the parameter ($\fenv=0.022$) no
thermal pulses were encountered. Instead the hydrogen-burning shell
corrosively burned into the H-free core and, eventually, into the CO core
\citep{Herwig2004,herwig:04a}. With $\fenv=0.014$, hydrogen ingestion and
efficient TDU with $\lambda>1$ are found frequently from the second
thermal pulse onwards (see Fig.~\ref{fig:lambda-fenv}).
A summary of this series of models is provided in Table~\ref{tab:modelprops-sj}
(first entry and middle 3 entries).

\subsubsection{Combined effect of \fenv\ and \fPDCZ}
\begin{figure}
  \centering
  \includegraphics[width=1.\linewidth,clip=True,trim= 0mm 0mm 0mm
  0mm]{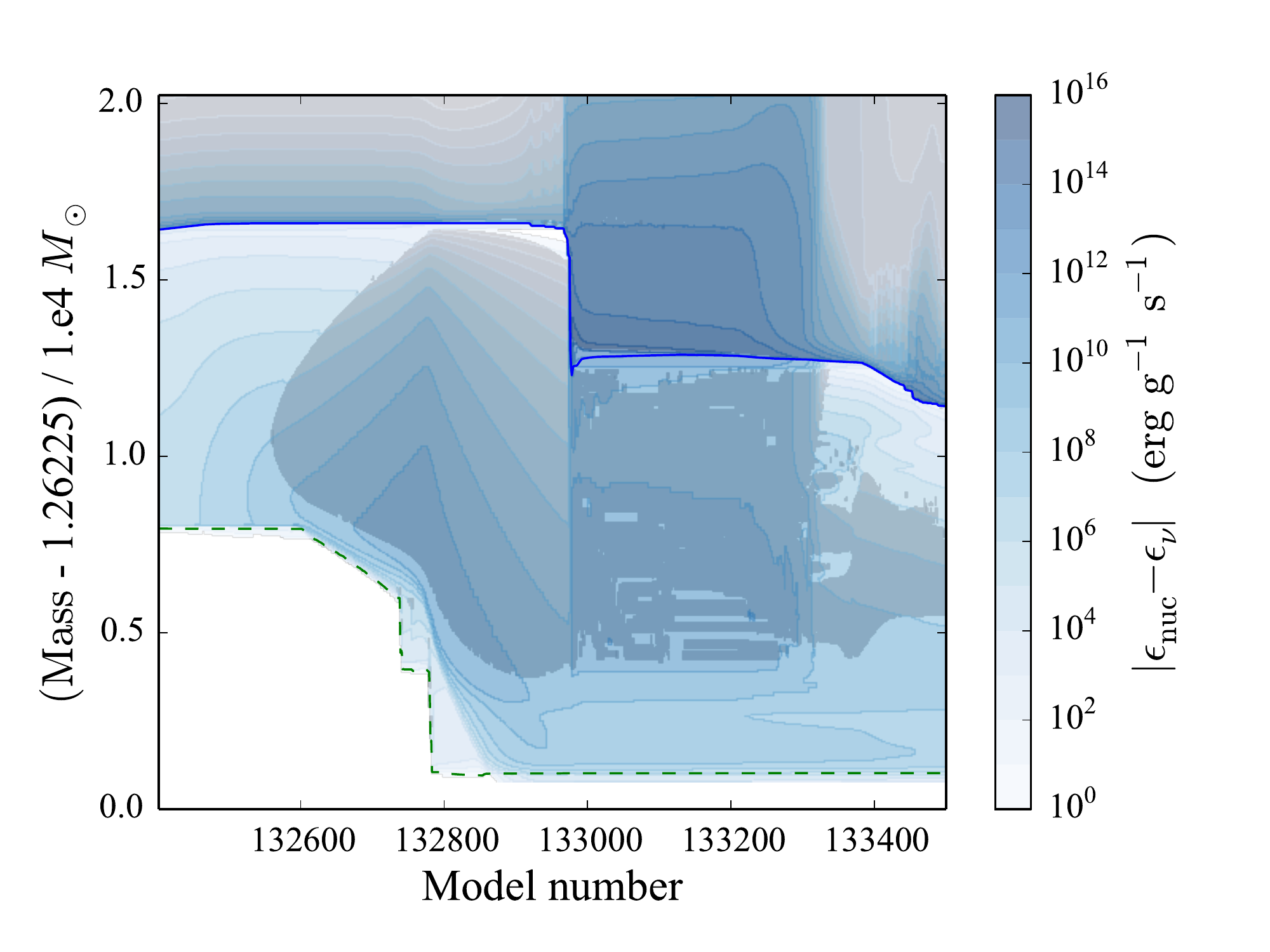}
  \caption{Evolution of the convective structure and nuclear energy generation during a thermal
  pulse in which hydrogen is ingested into the pulse-driven convection zone (PDCZ). The model
  shown here had initial mass $7\msun$ and metallicity $Z=10^{-4}$ and was computed with
  $\fenv=0.014$ and $\fPDCZ=0.014$ (penultimate model in Table~\ref{tab:modelprops-sj}). The regions shaded in
  grey are convectively unstable and the blue contours show regions of
  nuclear energy generation. The solid blue line marks the boundary above which the mass fraction of hydrogen
  is greater than $10^{-6}$ and the dashed green line marks the edge of the CO core. The ingestion
  event occurs around model number 133000, at which time profile plots of the important quantities are
  given in Fig.~\ref{fig:TP-ing-profiles}.}
  \label{fig:h-ing-kip-f-14-14}
\end{figure}

\begin{figure*}
  \centering
  \includegraphics[width=1.\linewidth,clip=True,trim= 7mm 0mm 7mm
  10mm]{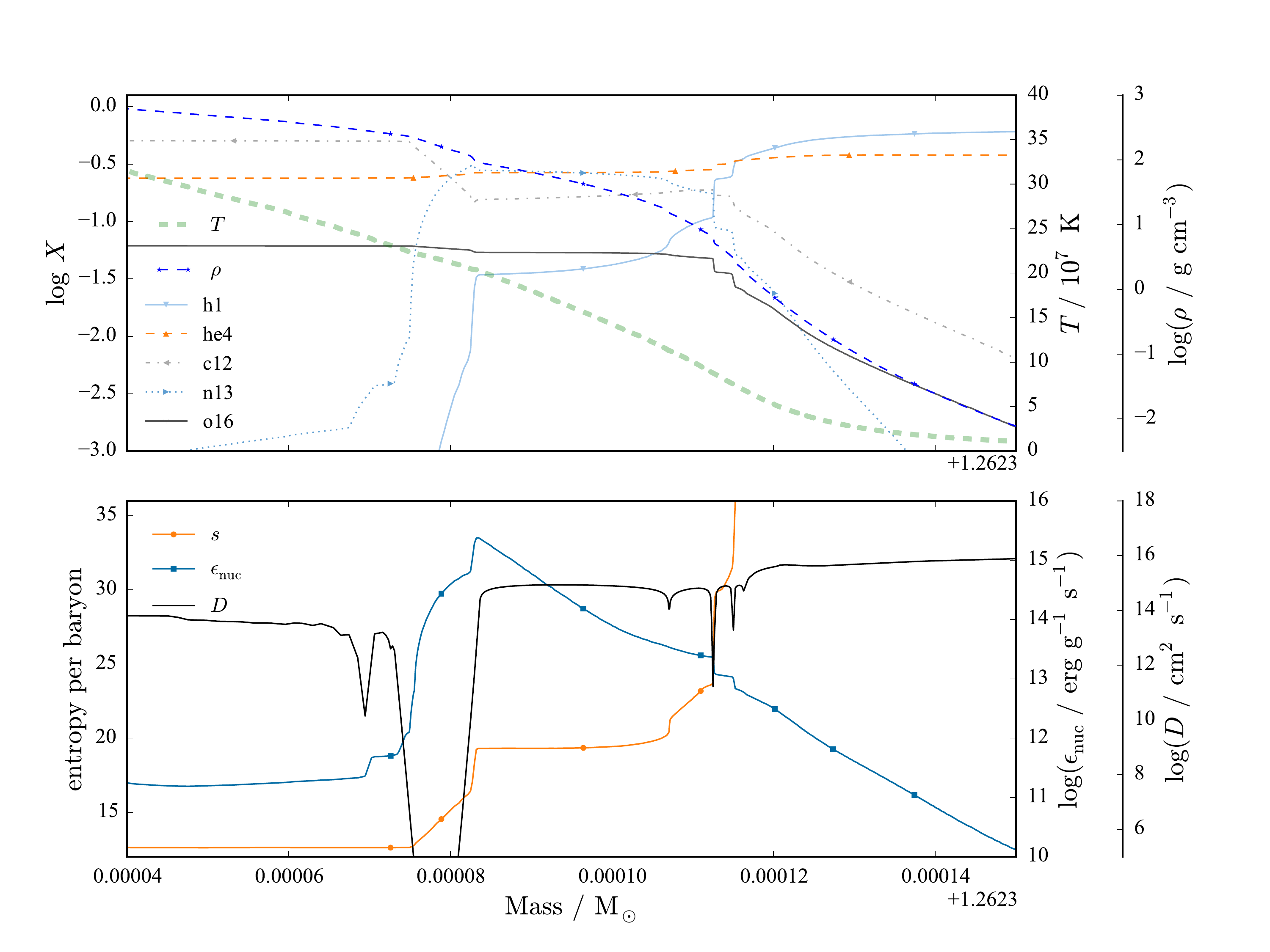}
  \caption{Abundance profiles, temperature and density (top panel) and entropy $s$,
  specific rate of nuclear energy generation and diffusion coefficient (bottom panel)
  in the 7\msun~$Z=10^{-4}$ model during the peak energy generation due to
  H-$^{12}$C combustion (around model 133000 in Fig.~\ref{fig:h-ing-kip-f-14-14}).
  In this model, the convective boundary mixing parameters at the
  bottom of the PDCZ and at the bottom of the envelope were $\fPDCZ=0.014$ and
  $\fenv=0.014$, respectively (see penultimate entry in Table~\ref{tab:modelprops-sj}.}
  \label{fig:TP-ing-profiles}
\end{figure*}

In a final series of models we increased both \fPDCZ\ and \fenv\ in tandem
(last 4 entries in Table~\ref{tab:modelprops-sj}).
The models behave as predicted based on our individual parameter studies for
each parameter. The combined effects of the CBM below the PDCZ and the
hot-bottom burning H envelope cause the model to experience H-ingestion TPs
very early in the TP-SAGB phase (within the first 20 TPs).
The model with $\fenv=0.014$ and $\fPDCZ=0.014$ experiences the highest
H-burning luminosity of all the models computed for this study:
$2.31\times10^{10}~\Lsun$.
Fig.~\ref{fig:h-ing-kip-f-14-14} shows a Kippenhahn (convective structure
evolution) diagram of this most extreme H-ingestion super-AGB thermal pulse
including contours of nuclear energy generation in the range
$1-10^{16}~\mathrm{erg~g^{-1}~s^{-1}}$ for the model with $\fPDCZ=0.014$ and
$\fenv=0.014$. The model predicts energy generation rates from hydrogen
combustion in excess of $10^{15}~\mathrm{erg~g^{-1}~s^{-1}}$.
The abundance profiles at this time are shown along with the temperature
(top panel) and entropy (bottom panel) stratifications in
Fig.~\ref{fig:TP-ing-profiles}.
The model was computed for only 14 thermal pulses, however the value of the
$H$-number (Eq.~\ref{eq:h-number}) during the second thermal pulse in this
case was $H\approx1.1$.
This suggests that the amount of energy being released in the ingestion event is
likely at least 10\% greater than the internal energy of the material residing
at the location where the energy is being released.
If the CBM parameter choice is appropriate then the 1D hydrostatic stellar
evolution assumptions are violated and the calculated results are
indicative at best (see Section~\ref{sec:consequences}).
For example, such strong energy release will certainly affect the hydrodynamic
flow which will feed back into the ingestion event.

\section{Consequences of hydrogen ingestion during dredge-out and the TP-SAGB}
\label{sec:consequences}

We have identified in Section~\ref{sec:results} two potential sites of
H-ingestion in super-AGB stars and failed massive stars. These are stars that
experience either dredge-out, He shell-flash thermal pulses or both.
In the dredge-out, the region of convective instability associated with the
He-burning shell grows outwards in mass and eventually coalesces with the
convective envelope. This allows for the transport of protons down to
He-burning temperatures on a convective timescale, where \carbon\ has been
produced from He burning. Introducing protons to an environment rich in
\carbon\ at temperatures of $\gtrsim1.5\times10^8$~K results in a very rapid
release of energy from H-\carbon\ combustion \citep[see also][]{Gil-Pons2010}.

Along the TP-SAGB, our models (assuming $\fPDCZ=0.002$ and $\fenv=0.0035$; see
Section~\ref{sec:method}) suggest that super-AGB stars with a wide range of initial metallicities
($10^{-5}\leq Z \leq 0.02$) evolve towards the conditions required for an exchange of material between
the pulse-driven convection zone (PDCZ) and the convective hydrogen envelope. Such a direction of
evolution involves: (i) the shortening of the time between the end of the PDCZ and the onset of the
third dredge-up (TDU) into the intershell and (ii) the increasing of the
efficiency of the TDU, $\lambda$.

\subsection{Validity of the 1D approximation}
\label{sec:1d_validity}
From our simulations of dredge-out in
TP-SAGB stars---which are spherically symmetric models in hydrostatic equilibrium computed
using the MESA code and holding the assumptions described in Section\,\ref{sec:method}---we
have calculated the $H$-number (Eq.\,\ref{eq:h-number})
for the time when the H-\carbon\ combustion is most energetic.
This number is an estimation of the
amount of energy released over one convective turnover time scale in units of the local internal
energy of the stellar material. During dredge-out in an 8.4~\msun\ stellar
model with an initial metallicity
of $Z=0.01$ we find a lower limit of $H=0.11$. During a thermal pulse of a
7~\msun\ super-AGB star
model with an initial metallicity of $Z=10^{-4}$ and convective boundary mixing
parameters
$\fPDCZ=0.002$ and $\fenv=0.014$ we estimate that $H\approx1.1$. 
Such conditions will certainly feed back into the hydrodynamic flow of the
stellar material, which in turn will feed back into the entrainment and burning
rates \citep{Herwig2014}.

As we describe in Section\,\ref{sec:results}, after some time in our 1D calculations the ingestion
episode is quenched. This means that the flow of protons from the H-rich convective
envelope into the convective He-burning layer is choked off by the formation of an entropy
barrier. The formation of such an entropy barrier arises because the protons burn as they are
diffused inwards, down toward higher temperatures.
The time scales of mixing, burning and nuclear energy generation are similar to one another.
Under these conditions, it is questionable whether the mixing length theory of convection (MLT) can
provide the correct solution for the transport of entropy and nuclear species. The behaviour of
H-ingestion events would be characterized by: how many protons are mixed into the \carbon-rich
layer, how far in are they transported and where is the energy from H-\carbon\ combustion
deposited. For this, 3D simulations are necessary \citep[see, e.g.,][for similar examples of H-ingestion
in low-metallicity AGB stars and post-AGB stars]{herwig:11,Stancliffe2011,Herwig2014}.

\subsection{Mass ejection and transients triggered by proton-ingestion}
\label{sec:mass_ejection}
The simulations by \citet{Herwig2014} suggest that H-ingestion into \carbon-rich He-shell convection can trigger a non-radial instability, the global oscillation of shell H-ingestion (GOSH). It is conceivable that a GOSH instability in
a violent proton-ingestion event in a super-AGB star could drive an outburst that rises
through the star and is ejected in a pre-supernova explosion. At the
shell, temperatures can reach in excess of $2\times10^{8}$~K.
These outbursts could have implications for multiple classes of
``supernova'' explosions, both as precursors of narrow-line supernovae
and, as possible in explanations of specific peculiar subclasses of
supernovae.  Although the thorough study needed to determine the 
viability of these outbursts in explaining these subsclasses is 
beyond the scope of this paper, in this section we conduct a first 
look at the transient properties of these outbursts.

First and foremost, the mass ejecta from these outbursts can produce
the circumstellar debris needed to explain type IIn
supernovae. Observations of type IIn supernovae are providing a
growing list of properties of the circumstellar medium surrounding
these supernova explosions
\citep[e.g.][]{Miller2010,Kankare2012,Fox2013,Ofek2014,Moriya2014,
  Taddia2013}. \citet{Ofek2014} argue that over half of all type IIn
supernova have precursors within 1/3~yr of the supernova explosion
with masses exceeding 0.001\msun.  \citet{Moriya2014} instead argue
that the IIn supernovae in their sample are fit by enhanced mass loss
rates of $10^{-3}-1$\msun~yr$^{-1}$ for $5-60$ years prior to the
explosion with total ejecta masses lying between
$0.01-10$\msun. \citet{Miller2010} argued that dense clumps exist as
far out as $1.7\times10^{16}$\,cm.  For most of these shell-burning
outbursts, the explosion occurs too long before collapse to produce an
observed signal in the supernova light-curve.  However, in some cases,
our models can produce the observed IIn CSM distributions.

\citet{Corsi2014} and \citet{Gal-Yam2014} have seen evidence of
outbursts at the few year time-frame even for supernova not of the
type IIn subclass, the primary difference being the ejecta is smaller
and the outburst happened longer prior to the supernova explosion.
These explosions may also be an indicator of helium-shell
proton-ingestion outbursts.  To predict the shape of the light curve
that could be produced during the ejection of material caused by a
proton ingestion event occurring deep within the star, we mimic the
outburst by assuming the hot material is ejected from the burning
layer.  As it rises through the star, it expands and cools
adiabatically prior to bursting out of the stellar surface.  We assume
that the drive from the burning layer only allows the ejecta to expand
laterally, and the volume of the ejecta increases as the square of its
position until it is ejected from the star.  Using the temperature at
the burning layer, and assuming adiabatic expansion, we can calculate
the thermal energy of the ejecta as it breaks out of the star.

\begin{figure}
  \centering
  \includegraphics[width=1.\linewidth,clip=True,trim= 0mm 0mm 0mm 0mm]{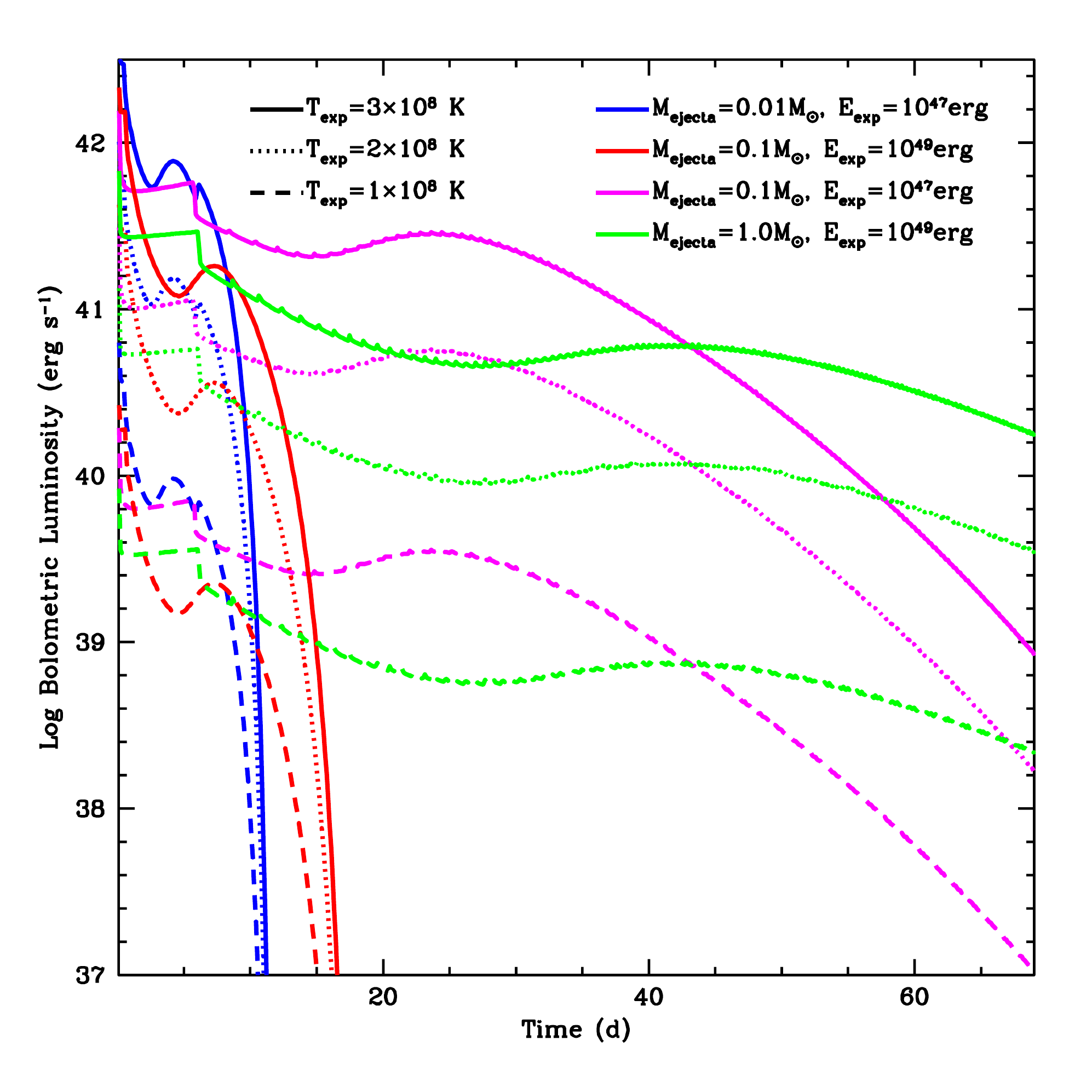}
  \caption{Bolometric light-curves of the possible signature from violent shell-burning explosions for
  different combinations of the shell burning temperature, ejected mass and the energy imparted to the
  ejected mass.}
  \label{fig:lightcurve_ejection}
\end{figure}

After the outburst breaks out of the star, we use a semi-analytic
solution to calculate the light-curve of these explosions, assuming
spherical symmetry and homologous outflow conditions \citep[see][for
  details]{Bayless2014}. To estimate the light-curve of these
outbursts, we varied the energy ($10^{47}-10^{49}\,{\rm erg}$),
temperature ($1-3\times10^8~\mathrm{K}$) and mass ($0.01-0.1 \msun$)
of this rising outburst.  Figure~\ref{fig:lightcurve_ejection} shows a
range of light-curves estimated from this method.  The calculations assume
average escape velocity $v_{\rm ejecta}$ set to $(2E_{\rm exp}/M_{\rm
  ejecta})^{1/2}$ where $M_{\rm ejecta}$ is the ejecta mass.  The
magnitude of the light-curve depends sensitively on the thermal
energy, but the duration of the burst can be altered by changing
either the explosion energy or the mass, and without an accurate
velocity measurement, it will be difficult to distinguish between
these models. 

Note that these outbursts lie somewhere between supernovae and novae and, for both small
ejecta masses ($<0.1\,M_\odot$) or energetic explosions, the duration is shorter than typical
supernovae with rise/decay times on order of 10 days.  
Without more 
detailed calculations of the propagation of the burning episode 
through the star, it is difficult to constrain the light-curve further.  
These light-curves may explain the pre-explosion outbursts seen 
by \citet{Corsi2014} and \citet{Gal-Yam2014}.
Similar pre-SN outbursts and CSM production could be caused by the instability
encountered in TP-SAGB stars described by \citet{Lau2012}, should the entire envelope not be ejected and the
star recover and evolve to an electron-capture supernova.
 
\subsection{Neutron-capture nucleosynthesis and overabundance of n-capture products in ejected winds}
H-\carbon\ combustion produces \carbon[13], from which neutrons are then released via the $\carbon[13](\alpha,n)\oxygen[16]$ reaction. Given the right conditions during the hydrogen-ingestion event (abundance of protons, abundance of \carbon\ and the temperature), intermediate neutron densities of $N_n\approx10^{15}$ cm$^{-3}$ can be produced, defined as the \iprn\ by \citet[][see Section \ref{s.introduction}]{cowan:77}.

\begin{figure}
  \includegraphics[width=1.\linewidth,clip=True,trim=11mm 3mm 15mm 13mm]{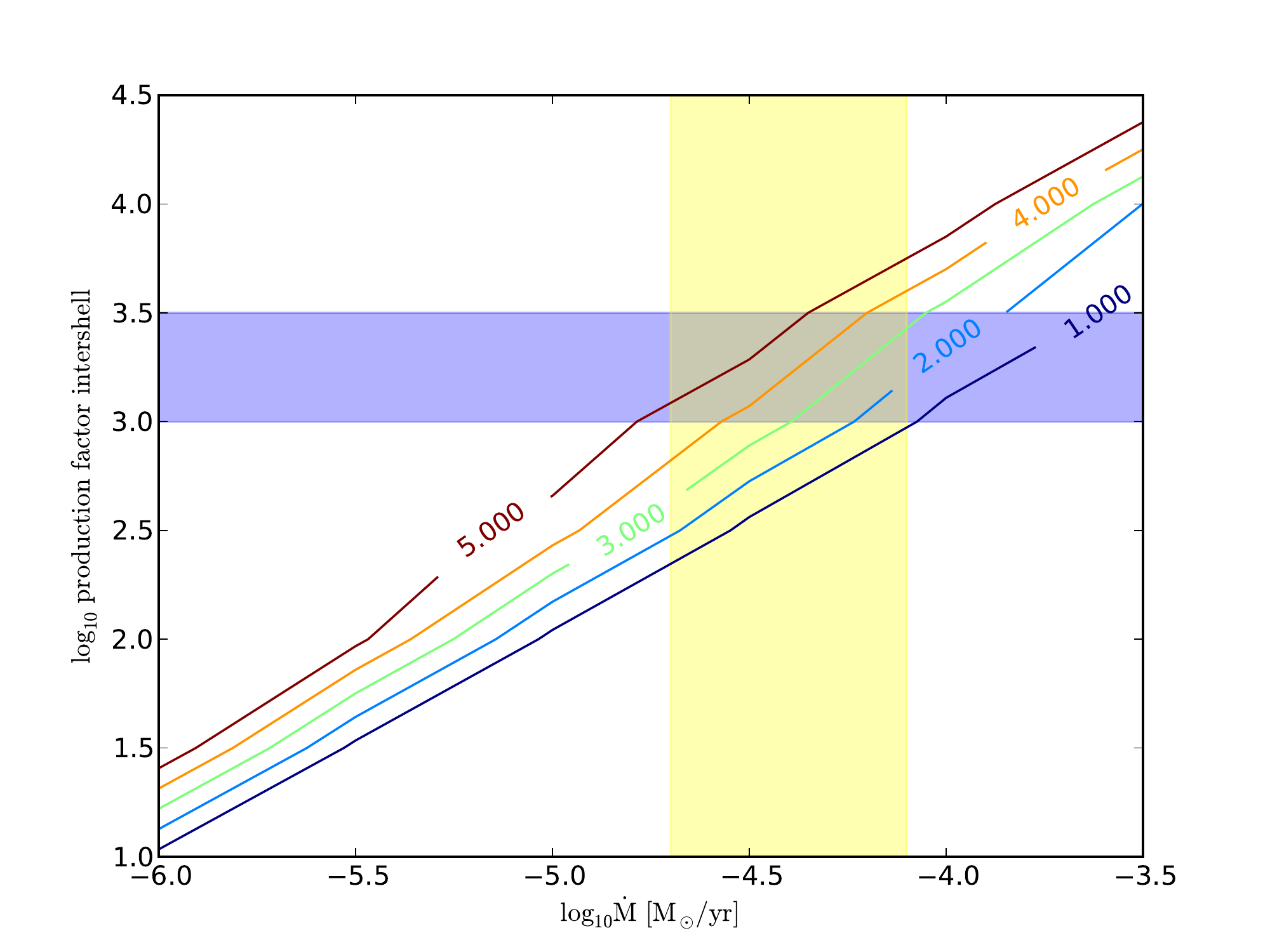}
    \caption{ Estimate of n-capture overabundance in ejected winds of
    super-AGB thermal pulse stars with H-ingestion flash, as a
    function of stellar mass loss and n-capture production factor in the
    intershell. The lines are labelled with $\log$ of the overabundance
    in the ejecta; if the intial heavy element abundance was solar
    scaled the labels would numerically correspond to [s/Fe] where s
    stands for the average of n-capture elements produced. The range
    of observed mass loss rates in luminous M-type giants, supposedly
    massive AGB stars \citep{vanloon:05b} is shaded yellow. The
    n-capture production factor estimates (shown in blue, see text) are based on
    the detailed nucleosynthetic H-ingestion flash investigation by
    \citet{herwig:11}, and our $\ipr$ single-zone calculations \citep{Dardelet2015}.
\label{fig:enrichment_estimate}}
\end{figure}

\ipr\ nucleosynthesis signatures are observed in young open clusters of the
galactic disk \citep{Mishenina2015} and may show up in galactic chemical evolution
at lower metallicities, or in metal-poor mass-transfer binaries
\citep{Dardelet2015,lugaro:12aipc,cristallo:09a}.
If low-metallicity super-AGB stars do indeed host the \iprn, then the
corresponding signature may be found in globular cluster stars with strongly
He-rich (second) generations \citep{Herwig2012mltalf,Charbonnel2013,
Karakas2014,Shingles2015}.

However, unless we have performed a three-dimensional hydrodynamic
simulation-based analysis of the interaction of nuclear burning and convective
fluid-dynamic flow, we can not calculate the nucleosynthesis in the
H-combustion dredge-up reliably based solely upon one-dimensional stellar
evolution models \citep{herwig:11}. In the meantime, we can explore a number
of relevant questions that are independent of the hydrodynamic
aspects. These are the enrichment of the envelope and ejecta of
super-AGB stars (which will be presented in this section) and the general properties of
nucleosynthesis at \ipr\ conditions. \citet{Dardelet2015} presented one-zone nucleosynthesis simulations for typical \ipr\ conditions, i.e.\ neutron densities of $N_\mathrm{n} \approx 10^{15}\mathrm{cm^{-1}}$. They have shown that the abundance pattern of some C-enhanced metal-poor stars classified as CEMP-s/r stars \citep[carrying simultaneously the signature of elements usually associated with \spr\ and those associated with \rpr;][]{Beers:2005kn,Masseron:2010hz,bisterzo:12,lugaro:12} can be remarkably well reproduced by one-zone simulations if the neutron exposure is used as a fitting parameter. If these preliminary results are confirmed in a more detailed investigation then H-ingestion events in super-AGB stars would be a possible site for the \ipr\ observed in CEMP-s/r stars. The C enhancement in CEMP stars would then likely come from the C-rich He-shell layer of which a larger fraction would be mixed  to the envelope and ejected during the violent non-radial instabilities associated with the H-ingestion flash \citep{Herwig2014} that are in contrast with 1D stellar evolution models. The predictions for \ipr\ simulations rely on nuclear physics data that is presently only available from theory \citep{bertolli:13}. Experiments for unstable species involved in \ipr\ simulations will have to be performed in the future to improve this situation. 

In order for \ipr\ nucleosynthesis from the SAGB hydrogen-ingestion thermal pulses
to become observable, the nuclear production in the
He-shell with H-ingestion needs to be high enough, and repeated events
are needed to enrich the envelope before it is lost to the interstellar medium.
One concern would be that the intershell in the super-AGB stars is so
small in mass that even repeated exposure and dredge-up events may not
lead to significant overabundances in the---at least initially---massive
super-AGB envelopes. One possible scenario mentioned above involves a dynamic ejection of envelope material triggered by the energy input and instabilities induced by the H-ingestion flash. Alternatively, in the TP-SAGB case the enrichment of the envelope may proceed more gradually, one dredge-up event at at time. 

Although the third dredge-up in some cases can be efficient in
terms of the dredge-up parameter ($\lambda \approx 1$; see Section~\ref{sec:CBM_impact}
and Figs.~\ref{fig:impact-fpdcz} and \ref{fig:lambda-fenv}) the mass of the
inter-shell region is only of the order $\approx 10^{-4}\msun$
(see Fig.~\ref{fig:h-ing-kip-f-14-14}). Even if large production factors can be
regularly obtained in the inter-shell region, the large dilution into a
massive super-AGB envelope may prevent large overabundance factors in
the ejected materials.

We have constructed a simple mixing
model to clarify this question.  In our simulation with $\fenv=0.014$
the core mass remains approximately constant due to
efficient dredge-up after each thermal pulse. Thus, the total number of
thermal pulses that the model could encounter would depend
only on the mass loss rate for a given envelope
mass. For $M_\mem{env}=6\msun$ and an approximately constant interpulse
time---because of the constant core mass with $\lambda \approx 1$---of
$t_\mathrm{interp} \approx 782~\mathrm{yr}$, a mass loss rate of $\log(\mdot
/ \msun\,\mathrm{yr}^{-1}) =-5$ ($-4$) translates into about 800 (about 80) thermal pulses.
The overproduction factor can be estimated by evaluating the following
simple mixing model:
\begin{eqnarray}
s_\mathrm{env}^\mathrm{n+1}=s_\mathrm{IS}^\mathrm{n}\frac{f_\mathrm{enr}m_\mathrm{IS}+m_\mathrm{env}^\mathrm{n}}{m_\mathrm{IS}+m_\mathrm{env}}\\
m_\mathrm{s, eject}^\mathrm{n+1}=m_\mathrm{s, eject}^\mathrm{n} +
s_\mathrm{env}^\mathrm{n+1} t_\mathrm{interp} \dot{M}\\
m_\mathrm{env}^\mathrm{n+1}=m_\mathrm{env}^\mathrm{n}-t_\mathrm{interp}
\mdot
\end{eqnarray}
where $s_{\rm env}$ and $s_{\rm IS}$ are the heavy element abundances in the envelope and
in the inter-shell respectively, $f_\mathrm{enr}$ is the production factor
of heavy elements in the inter-shell, $m_{\rm env}$ and $m_{\rm env}$ are the mass of the
envelope and the inter-shell respectively, $m_\mathrm{s, eject}$ is the mass of heavy
elements in the wind ejecta, and $\mdot$ is the mass loss rate.  The
overabundance in the total wind ejecta is then $m_\mathrm{s,
  eject}^\mathrm{final}/m_\mathrm{env}^{\mathrm{inital}}$, and
corresponds to [s/Fe] if s has the solar-scaled value in the initial
abundance. This overabundance is shown in
Fig.~\ref{fig:enrichment_estimate} as a function of assumed mass loss rate
and inter-shell production factors, assuming an average inter-shell mass
of $m_\mathrm{IS}=7\times10^{-5}\msun$ and an initial envelope mass of
$6\msun$.

Some guidance
on the expected production factors of the vigorous H-ingestion/He-shell flash
encountered in the super-AGB thermal pulses may be gained from the
related situation in post-AGB
H-ingestion flash models representing Sakurai's object
\citep{herwig:11}. There, production factors of about 100 were
found for an individual event, although limited to the first-peak
elements. 
The initial metallicity of our simulation with $\fenv=0.014$ is $Z=10^{-4}$
and therefore the production might be more efficient for heavier elements, since the neutron source is primary.
Consistent with the lower initial abundances, the logarithmic production factors in the inter-shell may be in
the range $3-3.5$ after each H ingestion event (see marked range in Fig.~\ref{fig:enrichment_estimate}).

We conclude that with an estimate of the super-AGB mass loss of $\log(
\mdot / \msun\,\mathrm{yr}^{-1}) \approx -4.1$ and production factors of
the order $1000$ in the intershell, we expect overabundances of the heavy \ipr\
elements in the wind ejecta of $1-2 \, \mathrm{dex}$. 
Therefore, the \ipr\ triggered by the $\carbon[13](\alpha,n)\oxygen[16]$ reaction may have a
comparable or even larger efficiency with respect to the \neon[22] neutron source activated
at the bottom of the convective TPs \citep{Doherty2014}.
On the other hand, the $\carbon[13](\alpha,n)\oxygen[16]$ reaction is a much more efficient
neutron source in producing heavier elements than $\neon[22](\alpha,n)\magnesium[25]$. Indeed, the
\neon[22] itself is also a relevant neutron poison and produces by the $(\alpha,n)$ channel another
neutron poison in \magnesium[25], reducing the neutron-capture efficiency beyond iron
\citep[see the behaviour of the $^{22}{\rm Ne}$ as a neutron source in fast rotating massive
stars and in low-mass AGB stars at low metallicity,][]{pignatari:13,husti:07}.

\section{Summary and Conclusions}
\label{sec:conclusions}

This paper describes two types of H-ingestion events arising in
simulations of 7--10\msun\ stars.
One type of event occurs in our super-AGB (SAGB) stellar evolution
simulations during the thermal pulse phase and the other occurs during the
dredge-out phase in SAGB stars and some stars that ignite neon off-centre.

Our simulations of SAGB stars with initial metallicities in the range
$10^{-5}\leq Z \leq 0.02$ include mixing processes at convective 
boundaries (CBM; \citealp{freytag:96,herwig:97}).
All SAGB star models evolve during the thermal pulse
phase toward conditions required for hydrogen-ingestion to occur during the He-shell
flash. We suggest that this evolution is characterised by two main
behaviours: (i) the efficiency of the penetration of the base of the convective envelope into
the He intershell (third dredge-up, TDU; see~\Sect{tp-evol}) increases as the stars
evolve along the thermal pulse phase and (ii) the TDU begins before the
pulse-driven convection zone (PDCZ) has completely extinguished.

Hydrogen-ingestion thermal pulses occur in our SAGB stellar models
for a large range of the CBM parameter $f$ representing the mixing efficiency
below the hydrogen envelope and below the PDCZ (\fenv\ and \fPDCZ, respectively).
In particular, using the value of $\fPDCZ=0.008$ which has been shown to reproduce
observational characteristics of post-AGB stars \citep{werner:06} and nova shell
flashes \citep{Denissenkov2013}, we find hydrogen-ingestion thermal pulses to occur
frequently for both $\fenv=0.0035$ and 0.014 (and for values in-between).
It is important to stress that both of these values of \fenv\ are much lower than
what is usually assumed for AGB stars \citep[see, e.g.,][]{Lugaro2003}. However,
with dredge-up being hot in super-AGB stars one would expect a greater entropy barrier
at the bottom of the convective hydrogen envelope than in low-mass AGB stars and thus
a lower efficiency of CBM. Quite what the efficiency of the CBM is and hence, what an
appropriate choice of the parameter $f$ should be for the bottom of the convective
envelope, is unclear at present. Our models suggest that for $\fenv\gtrsim0.014$
the efficiency of the 3DUP $\lambda$ will be greater than unity and thus the He-free core
will not grow during the TP-SAGB. This
scenario would therefore suggest a very low electron-capture supernova rate from
super-AGB stars \citep[see also][]{poelarends:08}.

The dredge-out phase in massive super-AGB stars and some stars that ignite neon off-centre
is another situation in which protons and \carbon\ are brought together under He-burning
temperatures on a convective turnover time scale \citep[see also][]{ritossa:99,siess:07,Doherty2015}.
We have shown the results of a simulated dredge-out episode that indeed depicts precisely
these conditions \citep[see also][]{Gil-Pons2010}. The behaviour is less sensitive to CBM at the bottom of the convective
hydrogen envelope in this scenario. Indeed, we have simulated the phenomenon by assuming
no CBM takes place at all at this boundary and still the dredge-out occurs (in fact, several previously
published SAGB stellar models depicting dredge-out do not include the effects of CBM, e.g. \citealp{siess:07}).

Our simulations are performed in spherical symmetry and employ the
mixing length theory of convection with time-dependent mixing treated as a diffusion process. Under these assumptions, the protons burn as they
diffuse inwards and form an entropy barrier that chokes off further transport of
hydrogen into the hot, \carbon-rich layers. Whether such a spherically symmetric entropy
barrier would form in a real star and whether it would completely inhibit the transport of protons
into the \carbon-rich layers is an open question. However, similar conditions have been found
in simulations of the very late thermal pulse (VLTP) in Sakurai's object
\citep{herwig:11}. In that case observations can be explained if it is
assumed that such a barrier formation and prohibition of proton transport
does \emph{not} occur.

The H-combustion events we find in our models
are vigorous, leading to significant mixing events that bring together protons and \carbon\ at
He-burning temperatures. Detailed nucleosynthesis simulations for a
related case, Sakurai's object, suggest that
substantial production of trans-iron elements would likely occur in such
H-ingestion events, and that the n-capture nucleosynthesis would proceed
at a neutron density inbetween that of the \sprn\ and \rprn\ \citep{herwig:11}.
The simulation
of these \ipr\ conditions for Sakurai's
object indicate a strong production of the first-peak elements, while Ba
and La are not efficiently made, leading to a simulated second to first
peak \spr-index ratio of [hs/ls]$\approx -1.8$, roughly in agreement
with observations. Post-AGB stars have only a very small amount of about $10^{-4}\msun$ of H-rich envelope material left. This justified mixing assumptions that effectively limited the neutron exposure so that the observed abundance pattern of a large enhancement of first-peak but not second-peak elements could be reproduced. 

In the H-ingestion thermal pulses of super-AGB star models with CBM an envelope with several solar masses of H-rich material remains. This, as well as the higher neutron to Fe seed ratio in low-$Z$ stars, and the possibility of recurrent H-ingestion events in SAGB thermal pulses imply that H-ingestion events in SAGB stars could be a site for \iprn\ with higher neutron exposure. This would lead to large enhancements of second-peak elements as observed in some of the CEMP-s/r stars. In fact, preliminary studies indicated that some CEMP-s/r stars are indeed very well reproduced by \ipr-nucleosynthesis \citep{Dardelet2015}. We therefore propose that \iprn\ in low-$Z$ SAGB stars is a potential origin of some CEMP-s/r stars.
Since hot bottom burning takes place in super-AGB stars, their envelopes are O-rich and
not C-rich. If H-ingestion thermal pulse nucleosynthesis in super-AGB stars are responsible
for a subset of the CEMP-s/r stars, the enrichment is probably coming from the ejection of
intershell material during a dynamic response to the convective-reactive H-ingestion flash, such
as a GOSH (see \citealp{Herwig2014} and Section~\ref{sec:mass_ejection}). Such outbursts would
produce faint transients lying somewhere between supernovae and novae, and may explain a fraction of
transient events that are observed shortly before a supernova explosion.


\section*{Acknowledgments}

This paper arose, in part, out of discussions early on with Don VandenBerg,
to whom the authors are grateful.


SJ is a fellow of the Alexander von Humboldt Foundation.
FH acknowledges funding through a Discovery Grant from NSERC.
MP acknowledges support from the "Lend\"ulet-2014" Programme of the Hungarian
Academy of Sciences (Hungary) and from SNF (Switzerland).
MP is also thankful for support from EuroGENESIS.
MGB's research is supported by the US Dept.~of Energy, Office of Nuclear Physics.
BP is supported by the NSF under grants PHY 11- 25915, AST 11-09174, and ACI 13-39581.

\bibliographystyle{mn2e_fixed}
\bibliography{references2}

\end{document}